\documentclass[conference,compsoc]{IEEEtran}

\PassOptionsToPackage{table}{xcolor}

\usepackage{tikz}
\usepackage{amsmath}
\usepackage{multicol}
\usepackage{cite}
\usepackage{csquotes}
\PassOptionsToPackage{hyphens}{url}
\usepackage{url}
\usepackage{ifthen}
\usepackage{amssymb}
\usepackage{colortbl}
\usepackage{svg}
\usepackage{enumitem}

\usepackage{pifont}
\usepackage{hyperref}
\hypersetup{
    colorlinks=true,
    citecolor=black,
    linkcolor=black,
    urlcolor=black
}
\usepackage{cleveref}

\usepackage[T1]{fontenc}
\usepackage{color}
\usepackage{graphicx}
\usepackage[caption=false,font=footnotesize,labelfont=sf,textfont=sf]{subfig}
\usepackage{xspace}
\usepackage{multirow}
\usepackage{makecell} %
\usepackage{enumitem} %

\usepackage{ulem}
\usepackage{listings}
\usepackage[export]{adjustbox}
\usepackage{multirow}
\usepackage{booktabs}

\usepackage{algpseudocode}
\usepackage{algorithm}

\usepackage{float}

\usepackage{pinyin-macros}

\ifdefined\isFinalized
\newcommand{\NewCommentType}[3]{}
\else
\newcommand{\NewCommentType}[3]{\expandafter\newcommand\csname #1\endcsname[1]{{\color{#2}{#3: ##1}} }}
\fi

\usepackage[override]{cmtt} %

\usepackage[utf8]{inputenc}
\usepackage{pgfplots}
\DeclareUnicodeCharacter{2212}{−}
\usepgfplotslibrary{groupplots,dateplot}
\usetikzlibrary{patterns,shapes.arrows}
\pgfplotsset{compat=newest}

\usepackage{balance}

\newcommand{\parhead}[1]{\medskip\Parhead{#1}}
\newcommand{\Parhead}[1]{\noindent\textbf{#1}\hskip 0.5em\relax}

\ifpdf
\pdftrailerid{}
\fi

\NewCommentType{todo}{red}{TODO}
\NewCommentType{dml}{violet}{dml}

\newlength{\byteboxdim}
\newlength{\byteboxwidth}
\newcommand{\byteboxpre}{}
\newcommand{\byteboxpost}{%
\renewcommand{\byteboxpre}{\hspace{0pt}}%
\kern-\fboxrule\relax%
\ignorespaces\nopagebreak%
}

\definecolor{frame}{gray}{0.40}       %
\definecolor{h} {RGB}{242,242,242} %
\definecolor{hc}{RGB}{242,242,242} %
\definecolor{lp}{RGB}{111,246,253} %
\definecolor{l} {RGB}{204,253,255} %
\definecolor{q} {RGB}{245,243,224} %
\definecolor{in}{RGB}{252,229,229} %
\definecolor{dg}{RGB}{220,254,206} %
\definecolor{b} {RGB}{247,243,211} %

\newcommand{\chinese}[1]{\begin{CJK}{UTF8}{gbsn}#1\end{CJK}}
\newcommand{\scanairportscount}{2,883}
\newcommand{\telegramairportscount}{758}
\newcommand{\totalairportscount}{3,431}
\newcommand{\surveyrespondents}{1,667}
\usepackage[most]{tcolorbox}
\crefname{appendix}{Appendix}{Appendices}

\newtcolorbox{takeawaybox}{
    colback=gray!8,
    colframe=gray!35,
    coltitle=black,
    boxrule=0.4pt,
    arc=2pt,
    left=5pt,
    right=5pt,
    top=4pt,
    bottom=4pt,
    before skip=6pt,
    after skip=6pt,
    fonttitle=\bfseries,
    title=Takeaway
}

\newboolean{anonymize-authors}
\setboolean{anonymize-authors}{false} 

\author{\IEEEauthorblockN{Rumaisa Habib$^*$\IEEEauthorrefmark{4},
Mingshi Wu$^*$\IEEEauthorrefmark{2},
Shiva Shahandeh\IEEEauthorrefmark{2}, 
Min Ni\IEEEauthorrefmark{2}, 
Eric Wustrow\IEEEauthorrefmark{3}, 
Zakir Durumeric\IEEEauthorrefmark{4}}\vspace{2pt}
\IEEEauthorblockA{\IEEEauthorrefmark{4}Stanford University\quad \IEEEauthorrefmark{2}GFW Report\quad \IEEEauthorrefmark{3}University of Colorado Boulder}
    \IEEEcompsocitemizethanks{
        \IEEEcompsocthanksitem[] $^*$Authors contributed equally.
    }
}

\begin{document}

\title{Understanding the ``Airport'' Censorship Circumvention Ecosystem in China}

\maketitle
\begin{abstract}
In China, a burgeoning underground market sells citizens subscription-based censorship circumvention proxies known as ``airports.'' We present the first systematic study of this ecosystem, combining user surveys, social media analysis, and active network measurements. We find that airports are by far the most popular off-the-shelf censorship circumvention tool in China, used by over half of our 1,667~survey respondents, who cite their ease of use, performance, and access to geo-restricted services like ChatGPT and Netflix. By scanning the Internet and scraping Telegram announcement channels, we identify 3,431 active airports built on a handful of open-source toolkits. We subscribe to 35 airports and characterize their performance, which often surpasses direct connections through the Great Firewall due to a distinctive multi-hop architecture. However, airports also pose new challenges and security risks: they accept payment through commercial services like Alipay, suffer frequent government takedowns, and are difficult for clients to configure optimally. Many airports also deploy their own distinct censorship policies. Airports are far more widely used than other circumvention tools from the academic literature, but introduce new forms of fragility and control, offering both lessons and opportunities for future circumvention research.

\end{abstract}

\section{Introduction}
\label{sec:introduction}

In China, an underground market sells netizens access to commercial network proxies---known as ``airports''
(\begin{CJK}{UTF8}{gbsn}机场\end{CJK}, \pinyin{ji1chang3})---for bypassing the Great Firewall (GFW). The name likely stems from Shadowsocks, a widely used proxy protocol with a paper-airplane logo~\cite{shadowsocks-github}: operating a fleet of proxy servers resembles running an airport. The term ``airport'' also serves as a code word on platforms like WeChat where terms like ``proxy'' and ``VPN'' are often filtered. Vendors advertise their airports via well-known announcement channels (e.g., on Telegram), sell metered proxy ``subscriptions'' through self-service web portals that accept Alipay and WeChat Pay, provide plug-and-play configuration for popular proxy clients like ClashX~\cite{clashx-official} and Shadowrocket~\cite{shadowrocket-appstore}, and compete on price, speed, and access to popular geo-restricted platforms like ChatGPT and Netflix\@.

In this work, we present the first comprehensive study of the airport ecosystem.
Despite their popularity, airports remain largely absent from the academic literature. This is a missed opportunity: airports emerged organically to meet user needs, evolved within a competitive environment, and can inform future system design. Conversely, the research community can help make one of the most widely used circumvention tools more resilient and safe for users.

We start by anonymously surveying 1,667~users in China who regularly bypass Internet censorship (\S\ref{sec:survey}). Despite many alternatives, we find that airports are the most popular off-the-shelf approach for accessing censored content: 55\% of respondents use airports with many citing their balanced ease-of-use for accessing blocked sites, speed, stability, and cost.  Through Internet scans for web-based subscription portals and manual analysis of public Telegram channels, we identify 3,431~active airports (\S\ref{sec:identify-airports}). We subscribe to 35~airports and characterize their architecture, performance, and service access. Airports are inexpensive (users report spending a median \$2.80/month) and are typically offered through tiered, byte-metered plans (\S\ref{subsec:results}).
Despite their low cost, airport proxies provide sufficient bandwidth for streaming video during peak hours, and, in our experiments, often outperformed direct connections to uncensored sites (\S\ref{sec:performance-evaluation}).

Airports achieve this impressive performance through a distinctive architecture that decouples ingress and egress endpoints (\S\ref{sec:architecture}) with ingress nodes inside China and egress nodes outside the Great Firewall, reportedly connected by leased International Ethernet Private Lines (IEPLs)~\cite{IEPL-CMI_IPLC,IEPL-CTAmericas2018,IEPL-CTAP2024,IEPL-Kxs2023,IEPL-Runtushare2025,IEPL-Tizi2021,IEPL-UnicomIEPL, xiaoji2023iplc, kxs2025intranet,V2EX2026IEPLAirportLinePull,NodeSeek2024ShenzhenHongKongIEPL,Ermaozi2025AirportLineType,OpenNetCN2026ClashAirportRoutes}. This approach bypasses GFW-induced congestion~\cite{Zhu2020a} and enables destination-based routing to geo-restricted sites. Many airports offer tiered egress nodes priced by access to high-demand services. While convenient, this architecture also introduces new risks: users provide PII to unknown operators, utilize government-monitored payment channels, and trust orchestration software with a history of vulnerabilities (\S\ref{sec:challenges-faced}). In addition, many operators impose their own censorship policies. We measured access to a subset of the Citizen Lab CN test websites~\cite{citizenlab_cn_testlist} and find that 
over half of the tested domains (198 out of 368) were blocked by at least one airport
(\S\ref{sec:self-censorship}). 
For instance, 12 out of the 15 airports we tested blocked access to Falun Gong sites, and nine blocked access to overseas news outlets.

Our results suggest that airports are now the foremost off-the-shelf method for bypassing the Great Firewall of China to access censored content due to their ease of use, performance, stability, and low cost. This success is driven by an architecture distinct from previously proposed solutions from the research community. Despite their popularity, airports introduce new privacy and security risks and shift censorship control to opaque commercial operators. We hope our analysis enables the next generation of censorship circumvention systems that both better meet the needs of users and protect their security and privacy.

\section{Related Work}
\label{sec:related-work}

The Great Firewall of China (GFW) is a state-operated Internet censorship and
surveillance system. It blocks access to restricted websites through
IP blocking~\cite{Winter2012a},
TLS SNI filtering~\cite{Chai2019a, Hoang2024a, Zohaib2025a},
DNS injection~\cite{Hoang2021a,Pearce2017b,Bhaskar2022a}, and
HTTP inspection~\cite{Rambert2021a,Knockel2017a}, and it detects circumvention
through passive traffic analysis~\cite{Wu2023a} and active
probing~\cite{Alice2020a}.
A substantial body of work has analyzed the
GFW~\cite{Hoang2024a,Tsai2024a, Filasto2012a, Raman2023b, Raman2020c,Niaki2020a}.
In response, many tools have been designed to evade censorship in China
~\cite{Dingledine2006a,Pu2024a,Zillien2024a,Bocovich2024a,Fifield2012a,Chi2024a,Bock2019a,Fifield2015a,psiphon3, wustrow2011telex,
lantern, meek} using proxies~\cite{Bocovich2024a,spec-shadowsocks,
Dingledine2006a, freegate, ultrasurf}, domain fronting~\cite{meek}, or
obfuscation~\cite{psiphon3, lantern, obfs4}. Several free tools operated by
volunteers outside China, including Tor~\cite{Dingledine2006a} and
Snowflake~\cite{Bocovich2024a}, enable users to bypass the GFW. Airports stand in 
contrast to these services: they are underground commercial operations run
from within China, with infrastructure both inside and outside the country.

Prior work has qualitatively examined censorship
circumvention~\cite{resisting-censorship-china, Feng2023a,
Xue2024b} and user perceptions~\cite{awareness-and-attitudes, resisting-censorship-china,
understanding-support, confucian,public-opinion-asia}. Xue et~al.~\cite{Xue2024a}
interviewed circumvention users and providers in China and Russia, documenting
the operation and challenges of commercial VPN and government-backed
circumvention services, and in subsequent work interviewed users and providers
to characterize their needs~\cite{Xue2024b}. Our work extends this line of
research by examining why users prefer airports and analyzing the architectural
and operational practices that let these services meet user demand.
Feng et~al.~\cite{Feng2023a} studied censorship circumvention in China in the
context of online gaming. 

A separate line of work has audited commercial VPN providers and consumer VPN
apps~\cite{khan2018empirical, ramesh2022vpnalyzer, ramesh2023all, Mixon-Baca2025a};
we treat airports as a Chinese variant of this commercial-proxy measurement
problem, one with distinct architectural and trust-model implications. Little prior work investigates commercial censorship-evasion tools, particularly the Chinese airport ecosystem, which faces a unique threat model and operational challenges. The closest is Chua et~al.~\cite{blackheart-airports},
who characterize the social dynamics of how users select airports on Telegram
and how the community polices scammers. Their work is complementary:
they study the community vetting of airports, while we analyze their popularity,
architecture, performance, and self-censorship.

\section{Airport Ecosystem Overview}
\label{sec:background}

\begin{figure}[t] \centering
	\includegraphics[width=\columnwidth]{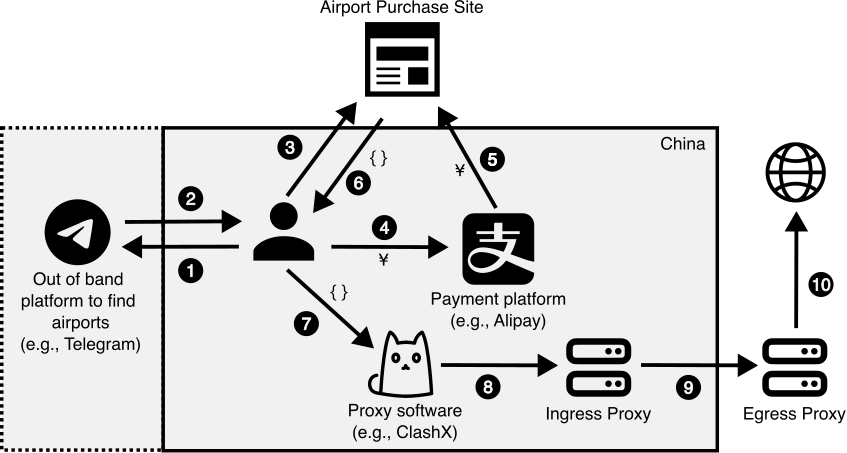}
	\caption{\textbf{Airport Walkthrough---}\textnormal{Users start by
	finding an airport through a third-party platform and creating an account
	(1--3), purchasing a subscription (4--5), and receiving a `subscription link'
	(6). The user then provides this link to a proxy client (7), which uses it
	to download and present the user with a selection of proxy nodes outside China to choose from. Through this software, the user connects to
	an ingress node in China (8), which connects to an egress node outside of China (9)
	to access the Internet (10).}} \label{fig:background}
\end{figure}

In this section, we provide an overview of the airport ecosystem, informed by
our investigation, online forum posts, and informal background conversations with
users, operators, and other researchers. Airports are commercial
proxy services for bypassing the Great Firewall in China. In this decentralized
ecosystem, several hundred individual operators within China run proxies and
sell metered ``subscriptions'' through online portals; these subscriptions can
then be imported into client applications and used to access the Internet freely.
We summarize the
process in~\autoref{fig:background} and detail each step below:

\parhead{Airport Discovery.} Because airports are run by independent operators,
users must first {locate} and {choose} an airport from which to buy service.
Within China, users typically find airports through: (1)~keyword
searches via domestic search engines (e.g., Baidu) or---using a traditional VPN---on Google;
(2)~direct recommendations from
peers; or (3)~access to foreign platforms like Telegram. For users who have
already partially bypassed the GFW, Telegram serves as a central hub where
operators pay well-known public channel administrators to promote their airports.

\begin{figure*}
    \centering
    \includegraphics[width=.8\linewidth]{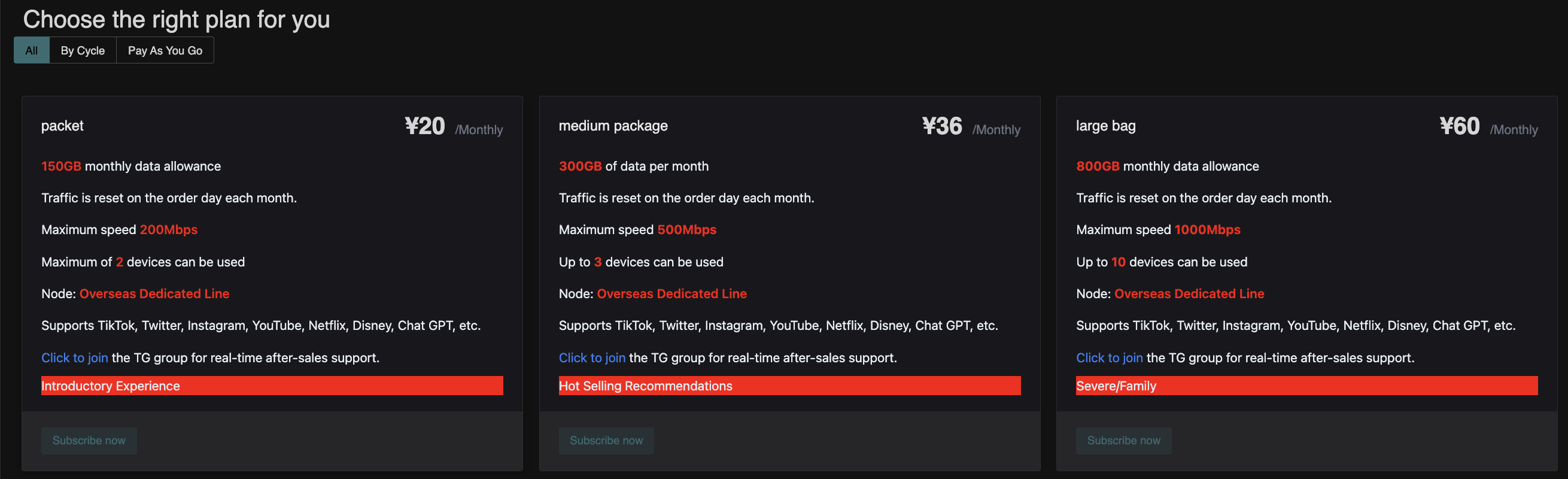}
    \caption{\textbf{Example Airport Subscription Plans}---%
    Airport portals present subscription tiers with usage quotas, durations, prices, and advertised service capabilities. In this example, the operator lists overseas dedicated lines and access to services commonly blocked in China, including Netflix and ChatGPT.
    }
    \label{fig:airport_sub_options}
\end{figure*}

\parhead{Portal Registration.} Airports provide web portals, largely built
from common open source templates, where users can
learn about an airport's services and purchase a subscription plan. In most
cases, users first register an account with a phone number or email, after
which they can view service options such as usage limits, service duration, proxy
node information, websites they enable access to, and cost. We show example plans from an airport in Figure~\ref{fig:airport_sub_options}.

\parhead{Subscription Purchase.} Users pay for access by purchasing  
\emph{subscription plans} that typically cost a few dollars per month, similar
to mobile data plans. Airports most commonly accept 
Alipay or WeChat Pay, though some support cryptocurrencies, PayPal, and
credit cards. To evade detection, operators often work with money-laundering
services to disguise revenue. For
example, one payment page advertises itself as \chinese{烟酒商行} (``Liquor \&  
Tobacco Store'').

\parhead{Subscription Links.} After payment, the portal provides users
with \emph{subscription links} that are compatible with client proxy software: each is a
URL pointing to a Base64-encoded JSON object that contains the configuration
to connect to the airport's proxy nodes. These links can be imported into
widely used
proxy clients such as Clash~\cite{clashverge-rev}, ClashX~\cite{clashx-official},
Shadowrocket~\cite{shadowrocket-appstore}, {sing-box}~\cite{singbox-github}, and Surge~\cite{ns-surge-official}).

\begin{figure}[t]
    \centering
        \includegraphics[width=\linewidth]{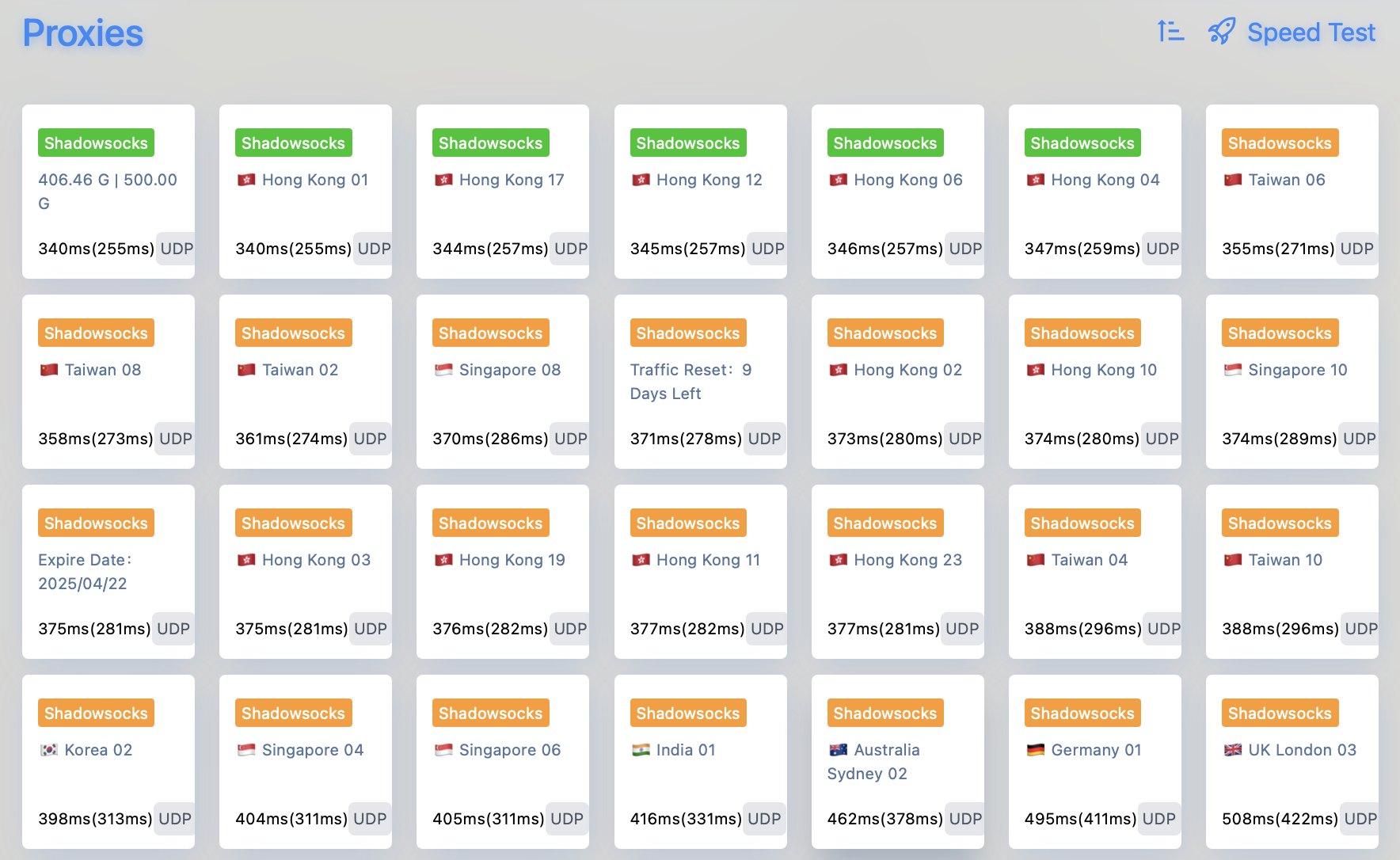}%
    \caption{\textbf{ClashX Proxy Selection UI---}Clients let users choose proxy nodes. Nodes with lower RTTs are green, indicating better performance.}
    \label{fig:client_ui}
\end{figure}

\parhead{Proxy Selection.} At launch, the user's chosen client fetches
the subscription link, parses the encoded configuration, and assembles a list of
available proxy nodes, which it presents to the user (Figure~\ref{fig:client_ui}). Most airports provide
several dozen nodes. Clients also display proxy metadata such as protocol, egress location, round-trip latency (via simple built-in benchmarks), and suggested usage 
(e.g., streaming-optimized or low-latency for gaming). Users can manually select a node 
or allow the client to auto-select based on latency measurements. Several clients support
complex routing rules and multi-node aggregation, enabling higher throughput by combining bandwidth from
multiple proxies.

\parhead{Monitoring and Renewal.} Throughout the life of a subscription,
the airport tracks the user's data consumption and
updates the subscription link (configuration file) accordingly; the client software
periodically fetches it to display key metrics, including remaining bandwidth,
latency, and subscription expiration. When a subscription nears
expiration or its quota is exhausted, the user can return to the airport portal to
purchase additional bandwidth or renew. This renewal process is
designed to be seamless, often involving one-click payment and automated link
regeneration.

\section{Why Users Choose Airports}
\label{sec:survey}

To understand the popularity of airports and why users select them, we surveyed over 1,600~Chinese Internet users who regularly employ censorship circumvention tools. We asked which tools they use, how they choose them, how much they pay, and what challenges they face. We find that airports are the most popular off-the-shelf mechanism for bypassing the GFW, used by 55\% of respondents. Users report choosing airports over other off-the-shelf tools for their greater stability and speed, and over self-hosting for their lower cost and reduced operational complexity.

\subsection{Survey Methodology}
\label{sec:survey-methodology}

We surveyed users in China by having a trusted organization within the censorship circumvention community (anonymized for submission) post a Qualtrics-based survey that we developed. Our survey seeks to understand:

\begin{enumerate}[topsep=0pt,itemsep=0ex,partopsep=1ex,parsep=0.5ex]
	\item How users in China bypass the GFW in practice;
	\item How users select their circumvention tool;
	\item How much users pay for circumvention services;
	\item What challenges users face with existing solutions.
\end{enumerate}

\noindent In this section, we focus on questions (1)--(3) to
understand airports' role and compare them to other circumvention methods.
We defer question~(4) to~\S\ref{sec:challenges-faced}.
Our study was approved by the Institutional Review Board (IRB) of the institution
that conducted the survey and analyzed the collected data.
We detail our ethical considerations in Appendix~\ref{sec:ethics}.
Below, we describe our survey instrument and recruitment.

\parhead{Consent.}
Before beginning the survey,
participants were required to acknowledge a consent form
(Appendix~\ref{appendix:consent-form}) that outlined the study's purpose,
our safety precautions, an overview of the research team,
and contact information for both the team and the overseeing IRB\@.
In particular,
participants were informed that the study concerned \textit{censorship in China}.
We did not offer compensation, given the survey's brevity (estimated five minutes to complete) and to avoid increasing risk for respondents.

\begin{figure*}[t]
    \centering

    \subfloat[Circumvention Method Popularity]{%
    \begin{minipage}[t]{0.31\textwidth}
        \centering
        \includegraphics[width=\linewidth]{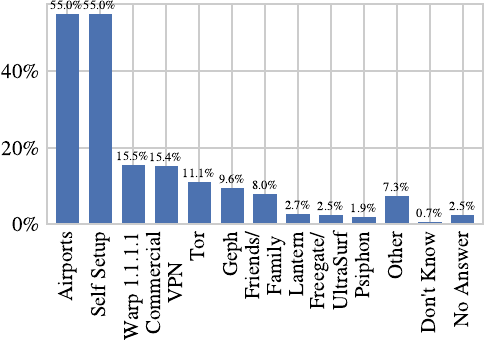}%
        \label{fig:survey-usage}
    \end{minipage}%
    }
    \hfill
    \subfloat[Selection Motivation By Method]{%
    \begin{minipage}[t]{0.31\textwidth}
        \centering
        \includegraphics[width=\linewidth]{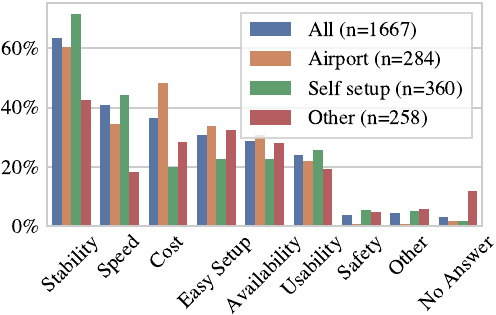}%
        \label{fig:survey-whyall}
    \end{minipage}%
    }
    \hfill
    \subfloat[Circumvention Cost]{%
    \begin{minipage}[t]{0.31\textwidth}
        \centering
        \includegraphics[width=\linewidth]{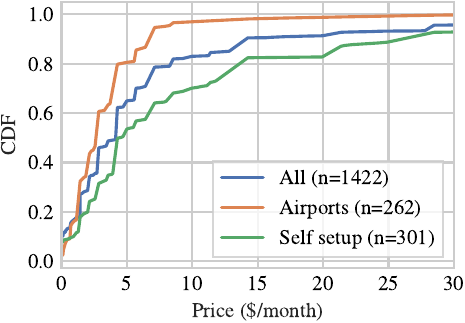}%
        \label{fig:survey-costall}
    \end{minipage}%
    }
    \caption{\textbf{Circumvention Preferences---}We anonymously surveyed 1,667~users in China to understand how they bypass the GFW\@. In (b) and (c), we only include respondents who chose a single method. Although safety was not a predefined survey option, 62~participants (3.7\%) mention it in free-form responses. Overall, stability is the most common selection factor, while airport-only users place more emphasis on cost than users of self-hosted setups.}
    \label{fig:survey-preferences}
\end{figure*}

\parhead{Survey Instrument.}
Our survey (Appendix~\ref{appendix:survey-questions}) includes six multiple-choice questions and one open-ended question on circumvention tool usage and preferences, as well as associated costs and challenges.
We did not collect personally identifying information,
and participants could skip any question or select \emph{Refuse to answer}.
For multiple-choice items, participants could select multiple answers and provide a free-form response.
Participants could take the survey in Simplified Chinese (default) or in English.
The survey was hosted on Qualtrics using a university-specific subdomain of \texttt{qualtrics.com}.
A bilingual author translated free-form responses from Chinese into English
and the team analyzed the responses; we did not perform back-translation or
compute inter-coder reliability for the free-form coding,
which we acknowledge as a limitation of the analysis.

\parhead{Recruitment.}
Because a survey about how users bypass government controls would likely be distrusted if posted by an unknown team,
we worked with a reputable party within the Chinese censorship circumvention community to post the survey on Twitter/X, Telegram, and the GitHub-based Net4People BBS
in October~2024~\cite{survey-x1}.
Recruitment posts were written in both Simplified Chinese and English,
with corresponding links to the survey in each language.

\parhead{Survey Responses.} Our survey ran for three months, from October~4, 2024 to January~10, 2025, yielding \surveyrespondents~valid responses out of 2,365 entries. We consider a response valid only if the participant provided explicit consent and reached 100\% completion before January~10, 2025. Of these, 1,612~(96.7\%) completed the survey in Simplified Chinese and 55~(3.3\%) in English. Most respondents reached the survey through links initially posted on Net4People (47.3\%) or the official Twitter/X account (41.2\%), and a smaller fraction through the official Telegram channel (6.5\%). These figures reflect the tagged entry links rather than the actual acquisition channels, as links were redistributed across platforms (e.g., Net4People links were later reshared in several Telegram channels).

\parhead{Limitations and Potential Sources of Bias.}
Studying censorship circumvention in China is difficult given the topic's sensitivity. Our survey has three main limitations. 

First, our recruitment introduces {selection bias}: because participation required encountering the survey through circumvention-community channels, respondents are likely more technically sophisticated, motivated, and engaged than the average circumventor in China. Our findings thus describe {active circumvention users} in these communities and do not necessarily generalize to the broader population. However, we do not expect any bias {toward} airport usage.

Second, responses are {self-reported} and may suffer from recall errors and social desirability or risk-related reporting bias, particularly as respondents may regard circumvention as sensitive; because we cannot independently verify each participant's location or circumstances, we focus on aggregate trends rather than claims requiring identity verification.

Third, our data is a {snapshot} from October~2024 to January~2025: individual airports churn frequently, so prevalence figures may have shifted by publication. The structural features we rely on---subscription-based commercial proxies, v2board/sspanel template dominance, and byte-metered pricing---have been documented across the ecosystem for several years and should remain stable over this window. 

Despite these limitations, our survey offers insight into a hard-to-study population: users who actively circumvent censorship and make recurring choices among off-the-shelf services, self-hosting, and free tools.

\subsection{Adoption, Choice Factors, and Spending}
\label{subsec:survey-results}

In this section, we present the popularity of airports and why users choose them over other circumvention tools.

\parhead{Tool Adoption.}
As shown in Figure~\ref{fig:survey-usage}, respondents prefer airports over other off-the-shelf solutions: 55\% of respondents use airports and 55\% use their own self-hosted proxy setup to bypass the GFW, followed by a long tail of other circumvention methods like Cloudflare Warp (15.5\%).

\parhead{Selection Priorities.}
As shown in Figure~\ref{fig:survey-whyall}, respondents choose tools based on stability (64\%), speed (41\%), and cost (37\%). Self-hosting users share these priorities but emphasize stability more (71\%). Airport users differ slightly: stability (61\%) remains the top concern, but cost (48\%) outranks speed (35\%). Respondents who use neither airports nor self-hosting prioritize stability (43\%), ease of setup (33\%), and cost (28\%). At a high level, users gravitate toward airports because they balance cost, speed, and stability. Self-hosted setups are faster and more stable but harder to operate ($z=2.6$, $p=0.004$)\footnote{Reported z-scores come from two-proportion z-tests, equivalent to a mean z-test on data encoded as 1 when an answer occurs and 0 otherwise.} and more expensive to maintain ($z=6.4$, $p\ll0.05$). Relative to other third-party tools, airports are more stable ($z=6.8$, $p\ll0.05$), faster ($z=8.1$, $p\ll0.05$), and less expensive ($z=5.3$, $p\ll0.05$).\looseness=-1

\parhead{Costs.}
Respondents pay \$0--8,119/month\footnote{Participants reported spending in Chinese Yuan (CNY), which we convert to USD at the October~3, 2024 exchange rate (1~USD = 7.02~CNY).} (median \$4.30) for their preferred tool (Figure~\ref{fig:survey-costall}). The wide range stems from a few outliers who run large-scale operations: we inspected all participants who paid over \$1,000/month and found that they all self-host, including one who pays several thousand dollars per month to lease a direct fiber link from China to Hong Kong. Respondents who use only airports report paying \$0--34/month (median \$2.80); some airports offer free plans, explaining the reported \$0 costs. By contrast, self-hosting respondents spend a median of \$4.60/month. This difference is consistent with cost being the second most common reason for choosing airports.

\begin{takeawaybox}
Airports are the most popular off-the-shelf circumvention option because they strike a unique balance of cost, usability, and stability. 
While stability is important across all circumvention tools, airport users emphasize cost more than other users; airport-only users report spending a median \$2.80/month. 
\end{takeawaybox}

\section{Airport Availability and Cost}
\label{sec:identify-airports}

We next investigate airport deployment and behavior, describing how we identify airports, the software powering them, and their high-level architecture.

\subsection{Discovering Airports}
\label{subsec:methodology}

We begin by analyzing public announcement channels on Telegram, which 53\% of our survey respondents reported using to find airports. In September~2024, we searched Telegram for the keyword ``\chinese{机场}'' (``airport'') and manually inspected the highest-ranked channels, identifying seven devoted to airport advertisements with over 5,000~members each (\autoref{tab:channels}). Manual analysis of hundreds of airports advertised in these channels surfaced two primary software platforms powering their sign-up portals: v2board and sspanel; informal conversations with members of the censorship circumvention community identified three additional platforms (\autoref{tab:templates}). We built fingerprints for each platform from its web resources (e.g., v2Board stores web resources in \texttt{/theme/v2board/assets/}, which we can identify).

\begin{table}[t]
    \centering
    \caption{\textbf{Public Airport Promotion Telegram Channels---}\textnormal{We extract public airport announcements from seven popular
    airport promotion Telegram channels as of September 20, 2024.}
    }
	\small
    \begin{tabular}{ l r@{\hspace{3pt}}r@{\hspace{3pt}}l r r}
        \toprule
        {Channel ID} & \multicolumn{3}{c}{{Established Date}} & {Members} & {Ads} \\
        \midrule
    @askahh        & Dec.& 2,  & 2019 & 16,670 & 1,501 \\
    @freemason6    & Apr.& 22, & 2020 & 35,126 & 618   \\
    @jichangtj     & May& 1,   & 2020 & 82,467 & 263   \\
    @cheap\_proxy  & May& 19,  & 2020 & 32,210 & 5,379 \\
    @jichang\_list & Feb.& 24, & 2022 & 36,503 & 3,468 \\
    @jctest6666    & Oct.& 5,  & 2022 & 7,611  & 497   \\
    @wxgqlfx       & Nov.& 2,  & 2022 & 23,935 & 882   \\
    \bottomrule
    \end{tabular}
    \label{tab:channels}
\end{table}

To find less widely announced airports, we scanned domains from 1,008~CZDS~\cite{czds} zonefiles, which provide coverage
over 84\% of the domains and 63\% of the TLDs seen in Telegram advertisements.
We used ZGrab~\cite{zgrab2} to make HTTP GET requests against domains on TCP/80 for fingerprinted
resources from an academic institution in November~2024. We followed the
best practices set by Durumeric et~al.~\cite{durumeric2013zmap,durumeric2024ten} when conducting scans. To ensure that we considered only active, working airports, we additionally instrumented Chromium using Playwright and filtered out URLs that did not return a successful HTTP 200 status code. We then manually investigated responsive websites and filtered out parked domains.

\subsection{Airport Subscription Portals}
\label{subsec:results}

We identified \totalairportscount~publicly accessible airport portals: \telegramairportscount~from Telegram and~\scanairportscount~through 
scans of known domains. This is not an estimate of the full ecosystem size, but a conservative floor for the number of active airports.

\begin{table}[t]
    \centering
    \caption{\textbf{Popular Portal Templates---}\textnormal{Airport portals typically use one of a handful of open source frameworks: v2board and sspanel account 95\% of the airports we find.
    The last column is not a summation because of overlap in identified airports across methods.}
	}
	\small
    \renewcommand{\arraystretch}{0.9}
\begin{tabular}{lrrr}
    \toprule
    Template & Telegram & Internet Scan & Total \\
    \midrule
    {v2board} & 424 & 1,896 & 2,175 \\
    {sspanel} & 191 & 974 & 1,100 \\
    default & 38 & 0 & 38 \\
    vendor & 11 & 0 & 11 \\
    xboard & 8 & 0 & 8 \\
    aurora & 7 & 0 & 7 \\
    {bob} & 0 & 6 & 6 \\
    {sspaneldjango} & 0 & 5 & 5 \\
    {zeropanel} & 0 & 2 & 2 \\
    Other & 79 & -- & 79 \\
    \midrule
    {Total} & {758} & {2,883} & {3,431} \\
    \bottomrule
\end{tabular}
\label{tab:templates}
\end{table}

\parhead{Software.} As shown in \autoref{tab:templates}, 95\% (3,275) of airport portals are powered by v2board and sspanel; the remainder use a variety of other packages, including v2board variants such as xboard. Both projects are widely used open-source platforms, with 4.7K and 9.6K~stars on their respective GitHub repositories as of April~2025. This standardization helps the ecosystem scale but also creates correlated risk: bugs, leaks, or misconfigurations in these open source projects can propagate across independent airports.

\parhead{Hosting.} About a third (27\%) of airport
portals geolocate to the United States per IPinfo~\cite{ipinfo} largely
due to the popularity of Cloudflare (82\% of U.S. airports). The next most popular location is Hong Kong
(3\%), followed by a long tail of other countries. The top three providers in
Hong Kong are Alibaba (28\%), DMIT (15\%), and Tencent (9\%).

\parhead{Limitations.} We cannot easily determine whether an individual airport is accessible at multiple domains or if an operator runs multiple airports because of common open source templates. However, we acknowledge that operators are motivated to operate an airport using multiple domains because of potential blocking. Informal background conversations with community members suggest that there are several hundred individual operators in China.

\subsection{Byte-Metered Pricing Model}

\begin{figure}[t]
    \centering
    \subfloat[All Plans]{%
    \begin{minipage}[t]{0.44\columnwidth}
        \centering
        \includegraphics[width=\linewidth]{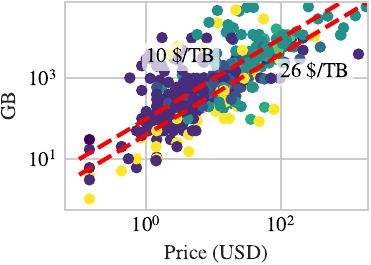}%
        \label{fig:sampled-price}
    \end{minipage}%
    }
    \hfill
    \subfloat[Least Expensive Plans]{%
    \begin{minipage}[t]{0.54\columnwidth}
        \centering
        \includegraphics[width=\linewidth]{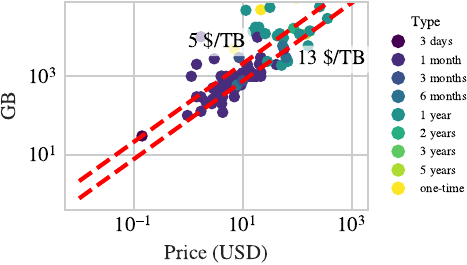}%
        \label{fig:best-price}
    \end{minipage}%
    }

    \caption{\textbf{Advertised Subscription Prices}---Panel~(a) shows price and data quota across all plans. Panel~(b) shows the least expensive \$/TB plan per airport. Price and data quota are shown on log scales. Red lines indicate the 25th and 75th percentiles. Data quota is the total traffic available over the full duration of the plan. Data cost varies significantly across airports and tiers, and longer plans tend to be more cost-efficient.}
    \label{fig:scrape-prices}
\end{figure}

Unlike traditional VPNs, which typically charge a flat fee~\cite{expressvpn-official,nordvpn-official,protonvpn-official}, airports adopt a byte-metered subscription model (e.g., 10~TB per month) similar to mobile data plans. In addition, some airports charge extra for bandwidth that transits specific proxy nodes (e.g., nodes that have access to premium services like ChatGPT). To characterize costs, we manually collected pricing information from 100~randomly sampled airports. As shown in~\autoref{fig:scrape-prices}, users pay a median \$16/TB, with substantial variation: plans range from \$0.17/TB to \$571/TB. 

Airports with longer subscription periods tend to be less expensive. Many airports also offer ``one-time'' tiers that provide a fixed data quota with no expiration; these are often the most expensive, costing a median \$29/TB. By contrast, longer-term subscription plans typically impose monthly caps despite charging for the full term: the median annual plan costs \$13/TB versus \$16/TB for a monthly plan. Monthly plans are the most common subscription duration, offered by 95\% of airports; the median monthly plan provides 500~GB for \$4.27, and the 25th-percentile plan costs \$2.56 for 150~GB\@. These figures are consistent with our survey, where the median respondent reported spending \$2.80/month (\S\ref{subsec:survey-results}), suggesting that most users select less expensive monthly plans.

To validate our measurements, we additionally examined eight popular airports recommended by members of the circumvention community in China. Prices vary widely even within this subset: the most cost-efficient plans range from \$0.97/TB to \$26.80/TB (\autoref{table:airport-information}). For context, a fixed 5~GB monthly broadband plan in China costs 0.42\% (\$4.28) of gross national income per capita (GNIpc)~\cite{ITU_ICT_Prices}; in comparison, extrapolating from the most cost-efficient \$/TB plans among popular airports, 5~GB through an airport averages 0.002\% (\$0.02) of GNIpc---roughly 0.5\% the per-byte cost of fixed-line broadband.

\subsection{Sampling for In-Depth Measurement}
\label{sec:sampling-airports}

To more deeply understand their behavior, we purchased subscriptions from a subset of the airports we found:

\begin{enumerate}
    \item Eight popular airports recommended to us by members of the censorship circumvention community in China;
    \item Nine less-popular airports found through our scans but never advertised on Telegram; 
    \item Nine airports that vary in age, as determined by when they were first advertised on Telegram: 3 old, 3 new, and 3 middle-aged airports; 
    \item Nine airports of varying price: 3~inexpensive, 3~expensive, and 3~medium priced.

\end{enumerate}

\noindent
In total, we purchased subscriptions from 35~airports (Appendix Table~\ref{table:airport-information}). For each, we subscribed to the least expensive plans that offered at least 200~GB `unrestricted' monthly traffic (i.e., plans that did not restrict traffic to specific services like Netflix or Facebook) using Alipay. Our subscriptions were blind to other features that may have been offered (e.g., dedicated IEPL lines or specific egress nodes). The subscriptions we purchased provided access to nodes that supported the Shadowsocks~\cite{shadowsocks} (55.41\%), VMess~\cite{vmess} (30.09\%), Trojan~\cite{trojan} (12.47\%), Hysteria~\cite{hysteria} (1.61\%), and Vless~\cite{vless} (0.42\%) protocols.
We describe our safety precautions when purchasing in Appendix~\ref{sec:ethics}.

\section{Performance}
\label{sec:performance-evaluation}
\begin{figure*}[t]
    \centering

    \subfloat[Round-Trip Time (RTT) Between Vantage Points and Ingress Proxy Nodes]{%
        \includegraphics[width=\linewidth]{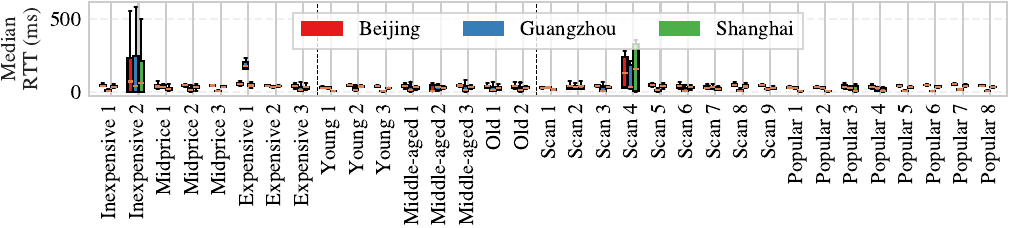}%
        \label{fig:all_rtt}
    }

    \vspace{0.5em}

    \subfloat[Time to First Byte (TTFB) to Download Cloudflare-hosted file Through Airport Node]{%
        \includegraphics[width=\linewidth]{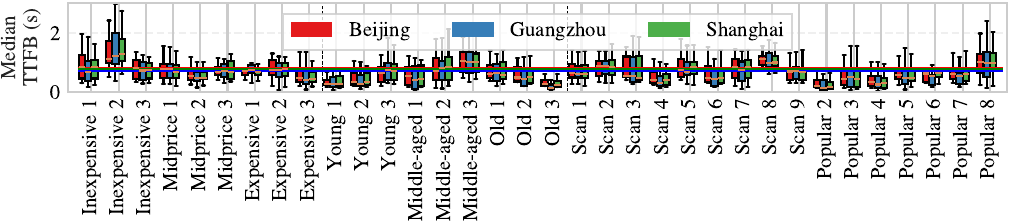}%
        \label{fig:all_ttfb}
    }

    \caption{\textbf{Airport Node Latency}---Individual datapoints represent the median time across a 24-hour period for a specific node from an airport in that location. Error bars extend to datapoints within 1.5$\times$ the interquartile range. The horizontal lines in (b) indicate the median baseline TTFB across a 24-hour period, measured from the three vantage points. While RTTs are similar, TTFB varies across categories: inexpensive, middle-aged, and unpopular airports tend to be slower.}
    \label{fig:speedtest}
\end{figure*}
\begin{figure*}
        \centering
        \includegraphics[width=\textwidth]{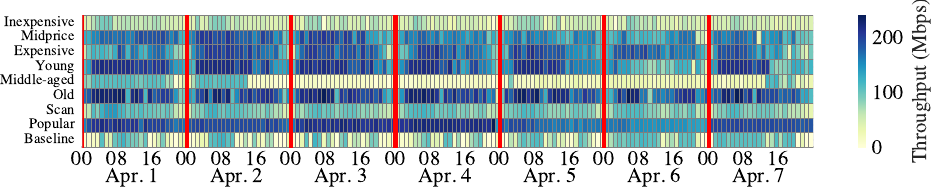}
        \vspace{-10pt}
        \caption{\textbf{Average Download Throughput---}\textnormal{Hour is given in Chinese Standard Time (UTC+8). Red lines indicate day boundaries.
		Throughput was averaged over all airports in their categories.
		Baseline indicates connections directly from the client to the server without using any proxies; both baseline and proxied flows terminate at Cloudflare PoPs.
		Using an airport proxy generally provides throughput comparable to or higher than the direct baseline for the destinations we test, plausibly reflecting that airport egress paths avoid the Great Bottleneck of China~\cite{Zhu2020a}, though we do not measure the underlying transit paths.
		In addition,
		even during the busiest hours, almost all airports offer over 5~Mbps of throughput, which is sufficient for streaming high-definition video. We see a similar pattern in Shanghai and Guangzhou, and omit the figures for brevity.}}
        \label{fig:diurnal_beijing_CDN_download}
\end{figure*}

We evaluate the performance of 35~representative airports by measuring 
download throughput, round-trip time (RTT), and time-to-first-byte (TTFB)
from three Alibaba Cloud instances in Beijing, Guangzhou, and Shanghai. Cloud instances had 2~cores and 8~GB RAM; none of the vantage instances reached maximum CPU utilization. Below we present measurements from Beijing; Guangzhou and Shanghai results were nearly equivalent.\looseness=-1

\parhead{Latency.}
We begin by analyzing latency because client software highlights latency to users to assist with node selection: low-latency nodes are most likely to be used. We characterize latency using two metrics: (1) round-trip time (RTT) to airports' {ingress} nodes (displayed to the client) and (2) time to first byte (TTFB) fetching a static 1KB file from Cloudflare (representative of experience). We conducted our measurements over a 24-hour period on March~29--30, 2025.

Most airports consistently provided multiple nodes with low latency between the client and ingress server, save for one inexpensive airport (Inexpensive~2) and one scanned airport (Scan~4), which were
particularly unstable (Figure~\ref{fig:all_rtt}).
Most airport--location pairs (e.g., Midprice--Shanghai) (80\%) had at least one node with a median RTT under 30~ms; five airport-location pairs (5\%) had at least one node with a median RTT of over 500~ms. 

The nodes with the lowest ingress RTT do not always have the best TTFB, which is more indicative of quality of
experience: 20 of 32~airports (63\%)\footnote{There were
32~airports for which we were able to get results for both our RTT and TTFB
experiments.} had differing fastest nodes depending on
metric. When considering TTFB (Figure~\ref{fig:all_ttfb}), inexpensive airports do not have significantly higher latency than expensive ones (0.94s vs.\ 0.90s). Purchasing a more expensive service does not necessarily provide a better experience.
Middle-aged airports are slower than both young ($t=3.8, p\ll0.05$) and old
($t=2.3,p=0.04$) airports.
Scanned/unpopular airports also provide a statistically slower experience than
popular airports ($t=2.2,p=0.04$), but not by much.
We report these tests as exploratory comparisons across small cohorts; effect sizes and per-cohort sample sizes
should be considered alongside the reported p-values.

Airports do not substantially increase latency relative to direct connections.
To test this, we conduct a baseline experiment by measuring the TTFB when downloading a 1\,KB file from the same vantage points but without an airport (indicated by the horizontal lines in Figure~\ref{fig:all_ttfb}).
The airports' average TTFB is comparable to that of the baseline
(0.9s vs.\ 0.8s). This difference in means is statistically significant under a t-test ($t=3.28$, $p=0.0011$), but is small in absolute terms.
Together, these measurements indicate that airports generally provide
usable node-level latency. 

\parhead{Throughput.}
We next measure download throughput, which captures whether airports can support high-bandwidth use cases like video streaming. Because bandwidth is metered by airports, we test only three proxy nodes per airport~\footnote{The breakdown of the protocols understood by these selected nodes are as follows: Shadowsocks (59.05\%), VMess (20.95\%), Trojan (18.10\%) and Hysteria (1.90\%)}. We specifically download data through three random nodes with RTT under 100~ms from clients in Beijing, Guangzhou, and Shanghai since user interfaces like Clash and Surge promote nodes with <100~ms latency as being ``green'', encouraging users to choose them. If fewer than three nodes meet this criterion, we select the three nodes with the lowest latency. Between April~1--7, 2025, we downloaded a 50\,MB file from Cloudflare every hour through our selected airports. We additionally measured a baseline comparison in which we directly downloaded the file without any airport.

As shown in~\autoref{fig:diurnal_beijing_CDN_download}, airport throughput is typically more than twice as fast as direct connections (median 131.60~Mbps vs.\ 64.24~Mbps) and exceeds the direct baseline in 72\% of observations. Throughput also remains consistent throughout the day: the median airport throughput in the lowest-throughput hour is still 71.65~Mbps (vs.\ 7.32~Mbps direct connection baseline). Most of the airports are consistently usable: 30 of 35~airports have median throughput above the 5~Mbps speed that Netflix recommends for full-HD streaming from 18:00--23:00 CST, when throughput is typically the lowest~\cite{netflix-speed-recc}.

While the median case is impressive, 15\% of observations were below 1~Mbps. These near-zero measurements are concentrated in a small number of airports (two middle-aged airports, one mid-priced airport, and one discovered through Internet scans), rather than being evenly distributed across the ecosystem. The maximum throughput we observed was 297.90~Mbps for an `Old' airport at noon CST.

There are significant differences across sampling categories. Popular airports have double the  
median throughput (208.66~Mbps) of other airports (104.32~Mbps). Intuitively this is consistent: airports are popular because they provide a better experience. Price also correlates with performance: inexpensive airports have substantially lower median throughput (45.82~Mbps) than both mid-priced (197.29~Mbps) and expensive (176.96~Mbps) airports. Lastly, expensive airports have the lowest variation, suggesting the most consistent performance. 

\parhead{Limitations.}
Our performance measurements have several limitations. First, because we measure throughput only on nodes with RTT under 100\,ms, our results likely reflect the more performant nodes that users tend to select rather than the full set of nodes in each airport. Second, our RTT and TTFB measurements span 24~hours and may miss day-of-week variation, while our throughput measurements span one week and may miss weekly or seasonal effects. Finally, we measure from three vantage points in China against a single external provider (Cloudflare); results from other parts of China or providers could differ.

\subsection{Decoupled Ingress and Egress}
\label{sec:architecture}

Airports have better throughput than direct connections to foreign websites likely due to their network architecture. Across the 35~airports we investigated, most ingress nodes geolocate to mainland China while most egress nodes were labeled as foreign. Ingress IPs are explicitly disclosed to users, but egress addresses are not. To recover egress IPs, we set up a custom IP-echoing HTTP server outside China and sent an HTTP GET request through each proxy node (\S\ref{sec:self-censorship}). We then geolocated the collected ingress and egress IPs using an IP2Location~\cite{ip2location} snapshot from April~8, 2025.

As shown in Tables~\ref{tab:ingress-egress} and~\ref{tab:city_level_ingress}, 92\% of ingress nodes geolocate to mainland China. Egress traffic, by contrast, nearly always exits outside mainland China---most commonly in Hong Kong, the United States, Japan, Taiwan, and Singapore. As subscribers, we observe ingress and egress nodes but not intermediate hops or the transport between them. Operators and community documentation report using dedicated cross-border links such as International Ethernet Private Lines (IEPLs) to bypass the GFW (Figure~\ref{fig:airport_sub_options})~\cite{IEPL-CMI_IPLC,IEPL-CTAmericas2018,IEPL-CTAP2024,IEPL-Kxs2023,IEPL-Runtushare2025,IEPL-Tizi2021,IEPL-UnicomIEPL, xiaoji2023iplc, kxs2025intranet,V2EX2026IEPLAirportLinePull,NodeSeek2024ShenzhenHongKongIEPL,Ermaozi2025AirportLineType,OpenNetCN2026ClashAirportRoutes}. While we cannot empirically confirm this architecture, it is consistent with the performance gains we observe, since IEPLs avoid GFW-induced congestion~\cite{Zhu2020a}.

Regardless of transport, architecturally separating ingress and egress nodes has several implications:

\begin{enumerate}
    \item Airports can use widely-supported protocols such as Shadowsocks and VMess for client access, while evolving ingress-to-egress transport over the GFW;

    \item Operators can replace egress IPs when they are blocked without requiring client reconfiguration;

    \item By concentrating cross-border transit into a managed segment, operators can invest in premium links where they matter most;

    \item Foreign egress nodes can be specialized for accessing services with geo-location or IP-reputation restrictions.

\end{enumerate}

\noindent Clients have also evolved to exploit this architecture in ways traditional VPNs do not. Client applications (e.g., Clash, Surge) and server software (e.g., Xray, V2Ray) support destination-based routing, sending traffic to different egress nodes depending on the destination. In practice, common airport clients let users import rule sets, assign proxy groups to specific destinations, and route domestic traffic directly while sending selected foreign or blocked traffic through airport nodes. This traffic splitting reduces quota consumption and improves performance on domestic services for users, while lowering server load and bandwidth use for operators.

\begin{takeawaybox}
    Most tested airports sustain usable cross-border throughput even during peak hours, often matching or exceeding direct connections. Their multi-hop architecture and destination-aware traffic splitting help explain this performance by separating domestic ingress from foreign egress and relaying cross-border traffic only as needed.
\end{takeawaybox}

\begin{table}[t]
    \footnotesize
    \caption{\textbf{Ingress and Egress Proxy Node Locations---}\textnormal{Ingress nodes are concentrated in mainland China (13~countries total), whereas egress nodes are distributed across other countries (76~countries).}}
    \centering
    \setlength{\tabcolsep}{3pt}
    {%
    \begin{minipage}[t]{0.33\columnwidth}
    \centering
    \begin{tabular}[t]{lr}
    \toprule
    \multicolumn{2}{l}{Ingress}\\
    \midrule
      CN   & 1,769 (91.9\%) \\
      TW   & 36    (1.9\%) \\
      SG   & 35    (1.8\%) \\
      US   & 29    (1.5\%) \\
      AU   & 15    (0.8\%) \\
      DE   & 12    (0.6\%) \\
      HK   & 8     (0.4\%) \\
      JP   & 4     (0.2\%) \\
      CZ   & 2     (0.1\%) \\
      VN   & 1     (0.1\%) \\
    \bottomrule
    \end{tabular}
    \end{minipage}%
    }
    \hfill
    {%
    \begin{minipage}[t]{0.30\columnwidth}
    \centering
    \begin{tabular}[t]{lr}
    \toprule
    \multicolumn{2}{l}{Egress}\\
    \midrule
        HK   & 239 (12.6\%) \\
        US   & 190 (10.0\%) \\
        JP   & 130 (6.9\%) \\
        TW   & 125 (6.6\%) \\
        SG   & 96  (5.1\%) \\
        GB   & 34  (1.8\%) \\
        KR   & 33  (1.7\%) \\
        DE   & 20  (1.1\%) \\
        TR   & 19  (1.0\%) \\
        AU   & 15  (0.8\%) \\
    \bottomrule
    \end{tabular}
    \end{minipage}%
    }
    \hfill
    {%
    \begin{minipage}[t]{0.27\columnwidth}
    \centering
    \begin{tabular}[t]{lr}
    \toprule
    \multicolumn{2}{l}{Egress contd.}\\
    \midrule
        RU   & 13  (0.7\%) \\
        FR   & 13  (0.7\%) \\
        CA   & 12  (0.6\%) \\
        VN   & 11  (0.6\%) \\
        AE   & 11  (0.6\%) \\
        IN   & 10  (0.5\%) \\
        NL   & 9   (0.5\%) \\
        AR   & 9   (0.5\%) \\
        MY   & 8   (0.4\%) \\
        IT   & 7   (0.4\%) \\
    \bottomrule
    \end{tabular}
    \end{minipage}%
    }
        
    \label{tab:ingress-egress}
\end{table}

\begin{table}[t]
    \centering
    \small
    \caption{\textbf{Geolocation of ingress nodes---}\textnormal{Ingress nodes are concentrated in mainland China, dominated by Guangdong and Beijing.}}
    \setlength{\tabcolsep}{2pt}
    {%
    \begin{minipage}[t]{0.42\columnwidth}
    \centering
    \begin{tabular}[t]{lr}
    \toprule
    \multicolumn{2}{l}{Location} \\
    \midrule
    China, Guangdong         & 636 \\
    China, Beijing           & 551 \\
    China, Shanghai          & 172 \\
    China, Hubei             & 133 \\
    China, Jiangsu           & 93  \\
    China, Gansu             & 80  \\
    Singapore, Singapore     & 35  \\
    China, Hainan            & 30  \\
    Taiwan, Taipei           & 29  \\
    USA, California          & 27  \\
    China, Anhui             & 26  \\
    \bottomrule
    \end{tabular}
    \end{minipage}%
    }
    \hfill
    {%
    \begin{minipage}[t]{0.54\columnwidth}
    \centering
    \begin{tabular}[t]{lr}
    \toprule
    \multicolumn{2}{l}{Location contd.} \\
    \midrule
    China, Fujian            & 21  \\
    China, Hunan             & 16  \\
    Australia, Queensland    & 15  \\
    Germany, Nordrhein-Westfalen & 11 \\
    China, Hong Kong         & 8   \\
    China, Shandong          & 5   \\
    Japan, Tokyo             & 4   \\
    China, Tianjin           & 3   \\
    China, Sichuan           & 3   \\
    Taiwan, Taoyuan          & 2   \\
    Czech Republic, Praha    & 2   \\
    \bottomrule
    \end{tabular}
    \end{minipage}%
    }
    \label{tab:city_level_ingress}
\end{table}

\section{Access Restrictions}
\label{sec:access}

Beyond performance, users also need to consider what sites airports enable accessing. We evaluate service access for airports and uncover airport-specific censorship policies.

\subsection{Bypass Server-Side Blocking}
\label{sec:server-side-unblocking}

First, we evaluate the extent to which airports can access high-value services, and compare airports against other well-known circumvention tools. 

\parhead{Methodology.}
From a Beijing VPS, we tested access to popular services using the open-source tool RegionRestrictionCheck~\cite{lmc9992025lmc999}, which fetches well-known sites and inspects the response for indicators of availability. We included services from the tool's multinational category and excluded tests that reported only the service's region of operation or conditions unrelated to availability (e.g., login or CAPTCHA requirements). We counted a service as available only when fully accessible, not when partially restricted (e.g., with content blocked upon VPN detection).

We evaluated service accessibility for 15 of the 35~airports we subscribed to that remained operational as of August 25, 2025. 
To contextualize these results, we also evaluated eight well-known tools using their official Android clients, enabling all available circumvention-related options (e.g., ``Shadowsocks Obfuscation'') when available.
Our goal is to measure service availability as a user would experience it through standard circumvention tools, not to isolate the effect of any single design choice, such as egress node selection or subscription tier. 

\newcommand{\noconnectrow}[1]{#1 & \multicolumn{10}{>{\columncolor{red!15}}c}{\textbf{Blocked by the GFW}} \\}
\newcommand{\goodcell}{\cellcolor{green!25}{\checkmark}}
\newcommand{\badcell}{\cellcolor{red!25}{\ding{55}}}
\newcommand{\grouprow}[1]{\rowcolor{gray!20}\multicolumn{11}{l}{\textbf{#1}}\\}

\begin{table}[t]
\centering
\caption{\textbf{Providers' Access to Online Services---}\textnormal{Airports generally enable access to more geo-restricted services than other circumvention tools. Green checks indicate full access; red crosses indicate blocking or partial availability. Amazon Prime Video (APV), ChatGPT (CGPT), Claude (Cld), Dazn (Dzn), Disney+ (D+), Google Gemini (GGem), Netflix (Nfx), Reddit (Rdt), TVBAnywhere+ (TVB), and YouTube Premium (YT).}}
\scriptsize
\setlength{\tabcolsep}{3pt}
\begin{tabular}{l cccccccccc}
\toprule
Tool & Dzn & D+ & Nfx & YT & APV & TVB & CGPT & GGem & Cld & Rdt \\
\midrule

\grouprow{Airport}
Airport~6   & \goodcell & \goodcell & \goodcell & \goodcell & \goodcell & \goodcell & \goodcell & \goodcell & \goodcell & \goodcell \\
Airport~7   & \goodcell & \goodcell & \goodcell & \goodcell & \goodcell & \goodcell & \goodcell & \goodcell & \goodcell & \goodcell \\
Airport~9   & \goodcell & \goodcell & \goodcell & \goodcell & \goodcell & \goodcell & \goodcell & \goodcell & \goodcell & \goodcell \\
Airport~10  & \goodcell & \goodcell & \goodcell & \goodcell & \goodcell & \goodcell & \goodcell & \goodcell & \goodcell & \goodcell \\
Airport~20  & \goodcell & \goodcell & \goodcell & \goodcell & \goodcell & \goodcell & \goodcell & \goodcell & \goodcell & \goodcell \\
Airport~8   & \goodcell & \goodcell & \goodcell & \goodcell & \goodcell & \goodcell & \goodcell & \badcell  & \goodcell & \goodcell \\
Airport~11  & \goodcell & \goodcell & \badcell  & \goodcell & \goodcell & \goodcell & \goodcell & \goodcell & \goodcell & \goodcell \\
Airport~23  & \goodcell & \badcell  & \badcell  & \goodcell & \goodcell & \goodcell & \goodcell & \goodcell & \goodcell & \goodcell \\
Airport~1   & \badcell  & \badcell  & \badcell  & \goodcell & \goodcell & \goodcell & \goodcell & \goodcell & \goodcell & \goodcell \\
Airport~3   & \badcell  & \badcell  & \badcell  & \goodcell & \goodcell & \goodcell & \goodcell & \goodcell & \goodcell & \goodcell \\
Airport~16  & \badcell  & \badcell  & \badcell  & \goodcell & \goodcell & \goodcell & \goodcell & \goodcell & \goodcell & \badcell \\
Airport~36  & \badcell  & \badcell  & \badcell  & \goodcell & \goodcell & \goodcell & \goodcell & \badcell & \goodcell & \badcell \\
Airport~37  & \badcell  & \badcell  & \badcell  & \goodcell & \goodcell & \goodcell & \goodcell & \badcell & \goodcell & \badcell \\
Airport~25  & \goodcell & \goodcell & \badcell  & \goodcell & \goodcell & \badcell & \badcell & \badcell & \badcell & \goodcell \\
Airport~26  & \goodcell & \goodcell & \badcell  & \badcell  & \goodcell & \goodcell & \badcell & \badcell & \badcell & \goodcell \\

\addlinespace[2pt]

\grouprow{Non-Airport}
NthLink~\cite{nthlink-official}    & \badcell & \badcell & \badcell & \goodcell & \goodcell & \goodcell & \goodcell & \badcell & \goodcell & \goodcell \\
Lantern~\cite{lantern}    & \badcell & \badcell & \badcell & \goodcell & \goodcell & \goodcell & \goodcell & \goodcell & \goodcell & \badcell \\
MullvadVPN~\cite{mullvad-official} & \badcell & \badcell & \badcell & \goodcell & \goodcell & \goodcell & \goodcell & \badcell & \goodcell & \badcell \\
\noconnectrow{Psiphon~\cite{psiphon3}}
\noconnectrow{Orbot~\cite{orbot-official}}
\noconnectrow{NordVPN~\cite{nordvpn-official}}
\noconnectrow{ProtonVPN~\cite{protonvpn-official}}
\noconnectrow{ExpressVPN~\cite{expressvpn-official}}
\bottomrule
\end{tabular}
 \label{tab:geoblock}
 
\end{table}

\parhead{Results.}
Of the eight non-airport tools tested,
only three functioned in China.
Among working tools, airports were able to access more services on average than non-airports (\autoref{tab:geoblock}).
Across the three working non-airport tools, 56.7\% of services were accessible, compared to 77.3\% for airports (on average). 
The difference is clearest for geo-restricted streaming services. All three working non-airport tools were blocked by Dazn, Disney+, and Netflix, while airports reached Dazn in ten of 15 cases, Disney+ in nine, and Netflix in six. Netflix was also commonly advertised by airport providers (Figure~\ref{fig:airport_sub_options}), suggesting that access to commercial media services is an explicit part of the value airports sell, even though Netflix remains the most restricted service in our test. In contrast, Amazon Prime Video was reachable through all tested airports and working non-airport tools.

Beyond streaming services, AI tools showed mixed results. Ten of 15 airports and one of the three working non-airport tools could access Google Gemini. Two airports were unable to reach ChatGPT, while all working non-airport tools were able to reach it.

\subsection{Self-Censorship}
\label{sec:self-censorship}

\begin{figure}[h]
    \includegraphics[width=\columnwidth]{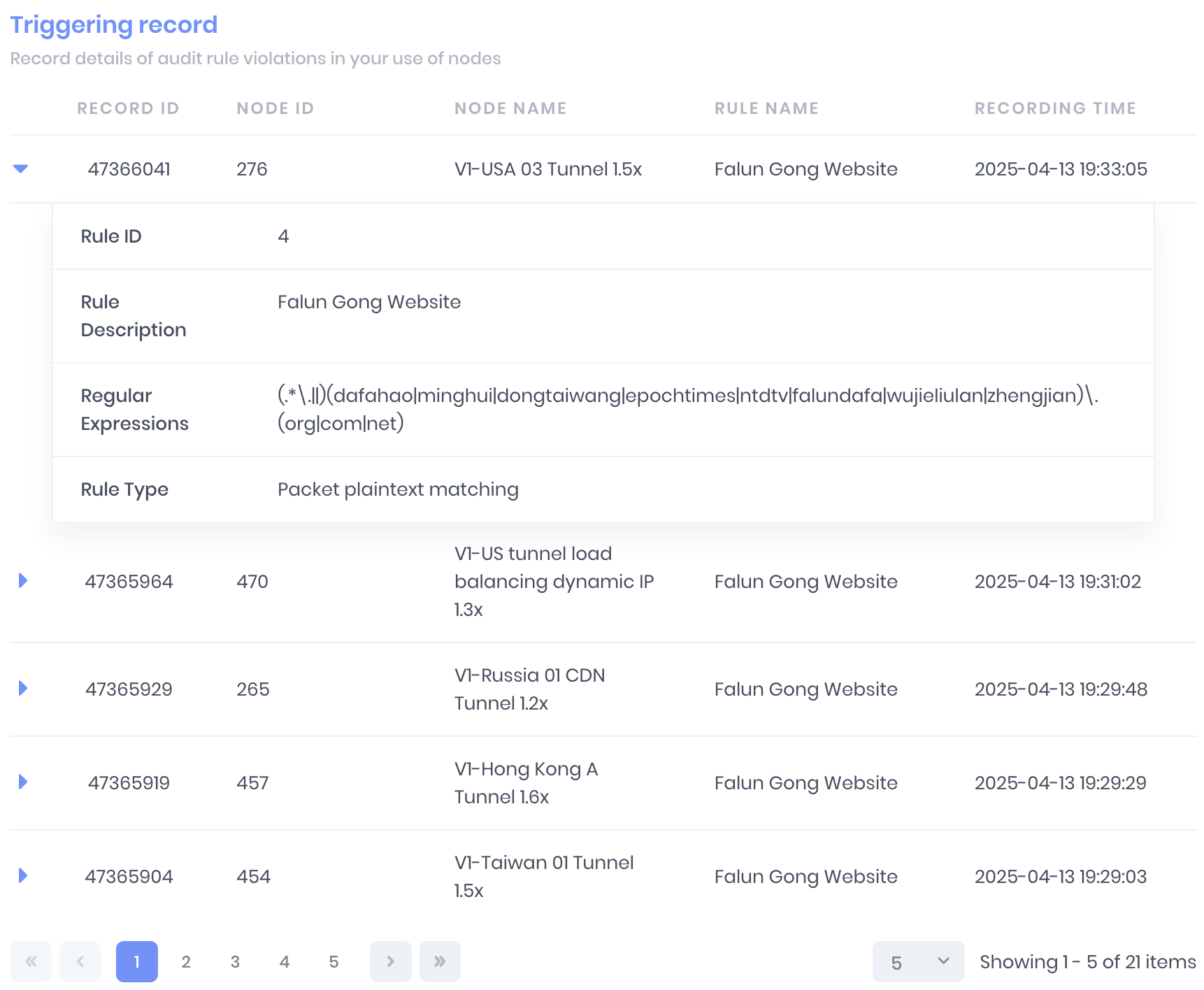}
    \caption{\textbf{Example Airport Violation Log---}%
    \textnormal{Each row corresponds to a specific rule violation,
    recording the timestamp, proxy node, and rule that was triggered.
    When expanded, the log provides additional details,
    including a description, the filtering patterns, and the matching method.}
    }
    \label{fig:self-censorship-log}
\end{figure}

Despite enabling users to access geoblocked sites and commonly censored content,
some users
report that airports block access to certain websites
(\S\ref{sec:challenges-faced}). In addition, some airports explicitly expose
audit logs that record requests that violate their access policies (\autoref{fig:self-censorship-log}). In this section, we measure the extent to which airports censor content and what motivations may explain this behavior.

\parhead{Methodology.}
We tested access to known censored domains through 15~airports---a subset of those from \S\ref{sec:performance-evaluation} that remained accessible as of August~23, 2025. (Individual airports are oftentimes ephemeral.) We began with 1,357~domains drawn from the Citizen Lab CN test list~\cite{citizenlab_cn_testlist} and from the audit rules and Terms of Service published by airports that disclose blocked sites~\cite{jichangtuijian2024,DuyaoSS_2018}. We narrowed this set to 368~domains that returned HTTP~200 over both HTTP and HTTPS from an uncensored U.S.\ network.
We then attempted to connect to each domain over HTTP and HTTPS through one randomly selected egress node per airport, from a client in Beijing (not all airports accept connections from outside China). We marked a domain as censored by an airport if both the proxied HTTP and HTTPS requests failed, since airports may filter on the HTTP Host header, the TLS SNI, or both. We excluded failures attributable to other causes (e.g., TLS misconfiguration), isolating cases consistent with deliberate blocking.

\begin{figure}[t]
    \centering
	\includegraphics{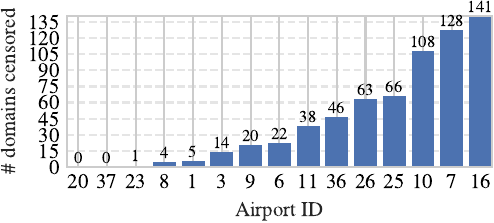}
	\caption{
        \textbf{Self-Censored Sites Per Airport---}%
        \textnormal{We tested 368~sites that were likely to be censored across 15~airports.
        Only two~airports did not self-censor any site.
        Over half (198~of 368) of sites were censored by at least one airport.}
    }
    \label{fig:num-censored-domains-by-airport}
\end{figure}

\parhead{Results.} Only two airports did not block any tested domain; 10 of the 15~airports censored at least ten domains (\autoref{fig:num-censored-domains-by-airport}). Of the 368~domains tested, 198 were censored by at least one airport. Using publicly available audit-rule categorizations~\cite{DuyaoSS_2018}, we manually classified these 198~self-censored domains into 16~categories (\autoref{tab:self-censorship-summary}); we describe the most prominent below. Together, these findings confirm that the user reports in~\S\ref{sec:challenges-faced} reflect a broader pattern: self-censorship is a systematic practice across airports.

\begin{table}[t]
\centering
\caption{\textbf{Top Censored Sites---}\textnormal{We categorized 368~blocked domains. ``Domain Sum'' is the number of domains blocked across all airports, while ``Domain Union'' is the number of unique blocked domains. ``Min,'' ``Med,'' and ``Max'' report the minimum, median, and maximum blocked domains in that category among airports that blocked at least one domain in the category. ``Airport Union'' reports the number of airports that blocked at least one domain in the category.}}
\begingroup
\setlength{\tabcolsep}{4.0pt}
\renewcommand{\arraystretch}{0.9}
\small
\begin{tabular}{@{}lrrrrrr@{}}
\toprule
\textbf{Category} & \multicolumn{5}{c}{\textbf{Domain}} & \multicolumn{1}{c}{\textbf{Airport}} \\
\cmidrule(lr){2-6}\cmidrule(l){7-7}
 & \textbf{Sum} & \textbf{Union} & \textbf{Min} & \textbf{Med} & \textbf{Max} & \textbf{Union} \\
\midrule
Falun Gong       & 307 & 57 & 1 & 19 & 56 & 12 \\
News \& Media    & 120 & 41 & 1 & 9  & 39 & 9  \\
Chinese website  & 84  & 42 & 1 & 7  & 31 & 8  \\
Other            & 31  & 12 & 1 & 9  & 10 & 5  \\
Trading          & 24  & 6  & 6 & 6  & 6  & 4  \\
Police           & 20  & 4  & 4 & 4  & 4  & 5  \\
Circumvention    & 15  & 5  & 5 & 5  & 5  & 3  \\
Tech             & 14  & 8  & 1 & 2  & 3  & 7  \\
Activism         & 8   & 8  & 1 & 2  & 5  & 3  \\
Finance          & 4   & 4  & 1 & 2  & 3  & 2  \\
\bottomrule
\end{tabular}
\endgroup

\label{tab:self-censorship-summary}
\end{table}

\parhead{Falun Gong.} Falun Gong-related websites were most commonly blocked: 12~airports blocked at least one domain, with 307~domain--airport blocks. Examples include minghui.org (the movement's official site), epochtimes.com (a Falun Gong-affiliated Western media outlet and known misinformation site~\cite{epoch_misinfo}), and wujieliulan.com (Ultrasurf~\cite{ultrasurf}, a circumvention tool developed by Falun Gong practitioners).

\parhead{News Media.} Nine~airports blocked access to news outlets.  Seven airports blocked {mingjingnews.com} (Mingjing News), five
blocked {chinadigitaltimes.net} (China Digital Times), four blocked {rfa.org}
(Radio Free Asia) and {voachinese.com} (Voice of America), and one blocked
{nytimes.com}. %

\parhead{Domestic Chinese Websites.} Eight airports blocked
access to popular Chinese websites and apps.  Notably, five airports
blocked the platform- and government-adjacent reporting portals {110.qq.com} and
{12321.cn}, which solicit reports of online ``harmful'' or illegal content.
Beyond these, the most frequently censored domains include security vendors
({360.cn}, {kingsoft.com}), online banking ({cmbchina.com}), media
({news.cctv.com}), and social platforms ({m.weibo.cn}, {xiaohongshu.com},
{douyin.com}). Airports likely block access to these domestic sites because they
could expose airports' egress IP addresses to companies and the government.

\parhead{Circumvention Services.} Three airports blocked
{torproject.org} (the Tor Project website), which prevents downloading Tor binaries.  This could be because Tor provides a free competitive offering.

\parhead{Inconsistent Enforcement Across Nodes.} In one instance, we find
that self-censorship is not applied consistently across all nodes for a given airport.
We performed a targeted probe using one of the
most censored domains ({epochtimes.com}).  For each of the 15~airports, we tested accessibility
across three randomly selected nodes.  Our results showed that at least one
airport (ID~6) enforced their policy inconsistently, permitting access on at least
one tested node while other nodes blocked the domain, indicating a potential
misconfiguration in this airport's blocklists.

\parhead{Self-censorship Motivation.}
Self-censorship may appear self-defeating since airports are bought to bypass censorship. We suspect that this is partly due to operator ideology. In one such incident, an airport operator was heavily criticized for suspending users who visited websites that the operator deemed `unpatriotic'~\cite{patriotic-airport-closure, telegram-aiguojichang}. The operator quotes a personal anecdote with Falun Gong as a reason for disabling access to sites that promote it. Another reason for blocking access to certain websites could be risk mitigation. While operating an
airport is illegal in China, blocking the most objectionable content
may lower the profile of operators, allowing them to evade targeting.

\begin{takeawaybox}
    Airports improve access to geo-restricted and censored services, but most tested airports also impose their own self-censorship, replacing state-imposed blocking with operator-imposed filtering. Our findings complicate the notion of ``free access": bypassing a censor does not guarantee a fully uncensored path. 
\end{takeawaybox}

\section{Ecosystem Challenges}
\label{sec:challenges}
Airports are popular because they make it easy to circumvent the GFW and access popular Internet services at a low cost. But this convenience also introduces fragility: users depend on opaque operators, third-party clients, and payment channels tied to individuals' identities. In this section, we examine challenges to and concerns about the airport ecosystem.

\subsection{User-Perceived Challenges}
\label{sec:challenges-faced}

\begin{figure*}[t]
    \centering
    \footnotesize
    \setlength{\tabcolsep}{6pt}

    \begin{tabular}{@{}c c@{}}
    \includegraphics[
        trim={0 0 0 0},
        clip,
        valign=c,
        width=2.5in
    ]{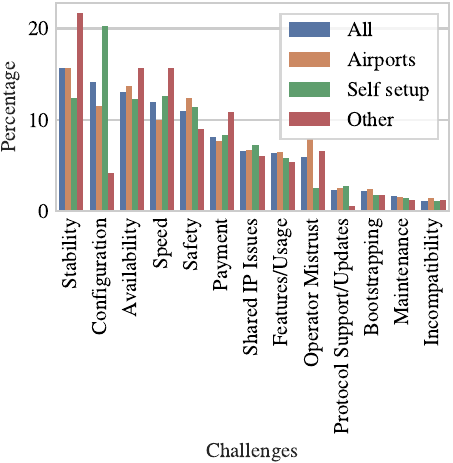}
    &
    \begin{tabular}{@{}>{\raggedright\arraybackslash}p{1in}
                    >{\raggedright\arraybackslash}p{2.7in}@{}}
        \toprule
        Challenge & Explanation \\
        \midrule
        Stability & Unreliable or frequently disconnected service \\
        Configuration & Difficulty setting up or configuring tools/software \\
        Availability & Inability to access the desired content, potentially because proxies or services are down, or because providers block specific websites \\
        Speed & Slow connections, high latency, or low bandwidth \\
        Safety & Fear of surveillance and/or legal consequences \\
        \midrule
        Payment Concerns & Risks from using payment methods such as Alipay or WeChat, or concerns about service price \\
        Shared IP Issues & Proxy IPs flagged by websites, leading to blocking or anti-bot mechanisms such as CAPTCHAs \\
        Operator Mistrust & Suspicion that airport operators log user traffic, cooperate with the government, or may disappear \\
        Limited Features/Usage & Missing features, such as auto-renewal or a friendly GUI, or usage limits such as device or traffic caps \\
        Protocol Support & Lack of support or updates for new protocols \\
        Bootstrapping & Difficulty acquiring or updating circumvention tools access requires bypassing the GFW \\
        Maintenance & Lack of development support or maintenance \\
        Incompatibility & Tools are unsupported on some operating systems, devices, or software configurations \\
        \bottomrule
    \end{tabular}
    \end{tabular}

    \vspace{1ex}
    \caption{\textbf{Reported Challenges---}We report the challenges most commonly faced by users in our survey.}
    \label{fig:survey-challengesall}
\end{figure*}

We draw on our survey responses to understand the challenges airport users face; 806~participants used the free-form response field to describe challenges and inconveniences (\autoref{fig:survey-challengesall}). Across all circumvention methods, users most commonly faced challenges with instability (180), configuration difficulty (163), and poor availability (150). For airport users, instability (16\%) was the most common challenge, followed by availability issues (14\%) whereas configuration was the largest challenge (20\%) for self-hosting respondents. Respondents who use neither airports nor self-hosting reported fewer configuration issues (4\%) but more stability issues (22\%) than both airport users ($z=1.86$, $p=0.03$) and self-hosting users ($z=3.03$, $p\ll0.05$), and reported speed problems more often (16\%) than airport users ($z=2.1$, $p=0.02$).

\parhead{Setup.} Users primarily found airports through Internet search
(62\%), Telegram (53\%), and friends/family (25\%). However, users note
technical challenges even before they begin using an airport: 17~respondents
noted the catch-22 that finding an airport service requires
circumventing the GFW: \chinese{想购买服务需翻墙，翻墙前需购买服务} (``Buying a service requires
circumvention, which creates a paradox.'').  Additionally, as airports often do not support
auto-renewal, users are left without access once
their subscription expires. This necessitates backup subscriptions, or fastidious 
manual renewals. Several respondents indicated the financial overhead of
keeping backups in case their current subscription ceased to operate. A respondent notes:
\begin{displayquote}
\chinese{运营不稳定：服务商的机器常面临被攻击、违规清退、ip被墙等问题，所以会同时购买多个机场的套餐以备不时之需} (``Operators’ machines often face attacks, abuse reports, and IP blocking issues, forcing users to maintain multiple subscriptions as backups.'')
\end{displayquote}
There are also technical challenges in configuring client proxy software. Participants noted that configuration is tedious, requires significant technical expertise, and has insufficient documentation or maintenance. This makes it difficult to share their tools with others. One respondent noted:

\begin{displayquote}
    \chinese{不容易帮助朋友设置 去年安卓手机客户端失效了, igniter停止了开发.}    (``Difficult to help friends set up; last year, Android clients stopped working, and Igniter stopped development.")
\end{displayquote}
Several respondents note the
difficulty of configuring traffic-splitting rules to ensure that only blocked
traffic goes through the airport nodes instead of all network traffic.
Additionally, users note that they need to be careful during configuration
to avoid DNS leaks.

\parhead{Operator Trust.}\quad Because airports are largely run by anonymous operators,
55~respondents (8\%) expressed mistrust, including concern
over potential data leaks
(\chinese{机场常常在未告知用户的情况下记录用户的网络日志} ``Airport services
often face data leaks; operators sometimes record network logs without informing
users.''). This fear may stem from airports that have previously been
compromised and had user data stolen~\cite{suyatu2022v2board_leak, planet-cx330_v2board-Data, bandwh2022v2board_leak, lomiweixiong2023v2board_breach}.
Some feared that the operator will cease to provide them their
services, whether through takedowns or by choice (\chinese{担心供应商跑路},
``Concern over providers abruptly shutting down.''). Participants also feared that
airport providers cooperate with the Chinese government. One
respondent reports:

\begin{displayquote}
\chinese{机场大多与中国政府有关系，安全性不太好} (``Most airports have ties to Chinese authorities, reducing security.'')
\end{displayquote}
\noindent Beyond general mistrust of airport operators, 47~users feared that they
or the operators would be caught by the authorities for bypassing the GFW\@. Airport users also expressed discontent with airport operators' political
statements and noted that ideological differences can deter potential airport
users who may prefer services that align with their own values. One user
expressed:

\begin{displayquote}
    \chinese{此外，有一个机场运营者的意识形态偏向“爱国爱党”，用大陆五星旗来标识台湾节点，我提出异议，希望至少去掉五星红旗，但没有被采纳。我怀疑他随时都会跑路。过去几年，台湾节点用五星旗的机场基本上都跑了。} (``one airport operator was ideologically ``pro-CCP'', marking Taiwan nodes with China's five-star red flag. I raised an objection, requesting at least the removal of the flag, but it wasn't accepted. I suspect they might go out of business at any time. Over the years, airports marking Taiwan nodes with the five-star flag have typically run away.'')
\end{displayquote}

\parhead{Censorship and Access.}\quad Ten respondents commented on the restrictive
self-censorship that some airports employ (described in~\S\ref{sec:self-censorship}).
For example, a user states:

\begin{displayquote}
    \chinese{我使用的Trojan机场订阅会对敏感网站（如.gov域名/境外新闻网站/时政论坛）进行封锁。} (``My Trojan airport subscription blocks sensitive websites like .gov domains, overseas news platforms, and political discussion forums.'')
\end{displayquote}

\noindent Additionally, airport proxy nodes often serve multiple users behind a single IP
address. As a result, 35~respondents report issues with websites classifying
them as bots and either blocking access or imposing CAPTCHAs. Moreover, the
content delivered to these users is localized to the proxy location, rather than
the user's, leading to a poorer quality of experience. For example, a user
reports:

\begin{displayquote}{}
    \chinese{日本线路下用Google 容易出现日文结果。} (``Using Japanese routes often leads to Japanese-language search results on Google.'')
\end{displayquote}

\parhead{Payment.}\quad Respondents note financial risks associated with 
airports, which mostly use Alipay or WeChat. Users fear
that airport providers may store their financial information, and express
concern over the privacy of the payment methods. While foreign payment
methods are safer, users note that they are inconvenient. Airports often offer
discounts if users purchase annual subscriptions. %
But users express that these longer subscriptions elevate
risk if there is a takedown or scam. One respondent mentions:

\begin{displayquote}
\chinese{服务商容易跑路，年付风险大但优惠大，月付季付相对麻烦没有优惠但风险小。} (``Operators often shut down suddenly. Annual payments carry risks but offer discounts; shorter-term payments are safer but less convenient.'')
\end{displayquote}

Some airport users (44) are also concerned about the cost of their subscription.
A user mentions: \chinese{年费相较于收入较高} (``Annual fees are relatively high
compared to income.''). This concern is further compounded by the fact that some
users (10) also feel that data plans are insufficient.

\subsection{Other Challenges}

Beyond the concerns expressed by surveyed users, our investigation also uncovers several systemic weaknesses:

\parhead{Vulnerability Impact.}
Airports have seen multiple data-leak incidents,
including disclosures affecting users of v2board-based portals~\cite{planet-cx330_v2board-Data,suyatu2022v2board_leak,bandwh2022v2board_leak,lomiweixiong2023v2board_breach}.
Because 95\% of the airports we identify run v2board or sspanel
(\autoref{tab:templates}), faulty code in a shared template can create vulnerabilities in 
hundreds of otherwise independent airports.
The same template standardization that lowers the barrier of entry to operate an airport
also makes it a correlated-risk surface.

\parhead{Vulnerability Disclosure Channel.}
Because airport operators are fully anonymous, there is no established channel for disclosing vulnerabilities, nor an easy way for users to learn whether their operator
has applied a fix. The closest replacement to a systematic vulnerability disclosure platform are informal, operator-run Telegram channels and announcements on individual airport portals.

\begin{takeawaybox}
Airport users commonly report instability and availability problems, showing that commercialized circumvention improves usability but still leaves users dependent on fragile services and opaque operators. Future tools should borrow airports' usability model, but pair it with stronger transparency, failover, and trust mechanisms.
\end{takeawaybox}

\section{Discussion and Conclusion}
\label{sec:discussion}
In this work, we systematically analyzed the airport proxy ecosystem in China, showing that airports have become a leading off-the-shelf censorship circumvention option due to their usability, performance, low cost, and ability to access geo-restricted services. At the same time, we uncover self-censorship and other challenges faced by users and operators. In this section, we draw lessons from airports' success and discuss recommendations for the censorship circumvention community.

\parhead{Decentralization Enables Resilience.} 
Airports are not technically sophisticated nor are they hidden. As a result, they are regularly blocked and taken down by authorities~\cite{MuchuanProcuratorate2017, DuiHua2019VPNPartII,ye2024seizing_innocence,zhou2024acquittal,zhou2022nong_case,tan2026chinafears}. Yet, despite the fact that individual airports are relatively fragile and ephemeral, the ecosystem appears resilient because of the large number of independent operators and proxy nodes available at any given time. That resilience is not a
product of protocol design. Rather, it comes from a handful of
open-source toolkits that make it easy and accessible to
run an airport. As a result, hundreds of people inside China are willing to run 
airports.

\parhead{Circumvention as a Service.}
Our results suggest that one of the foremost use cases for airports is accessing popular geoblocked and/or GFW-blocked services like Netflix and ChatGPT on a daily basis rather than viewing overtly politically sensitive content. Beyond ease of use and performance, this may explain why users are willing to pay for airports even though free circumvention tools exist. It may be that access to more explicitly censored political content is a side-effect rather than the primary goal of airports. Nonetheless, through their focus on mainstream users and websites, airports may also have unintentionally become one of the easiest ways to access other types of censored content. 

\parhead{New Security and Privacy Risks.} While there is a clear willingness to pay for access and payments enable the airport ecosystem to be self-sustaining, commonly used payment processors like Alipay and WeChat create clear risks to users, who can be easily identified. This creates a new challenge for future systems: how to accept payment in an easily accessible manner that simultaneously provides user security in an ecosystem where cryptocurrencies and other payment methods are banned.

\parhead{Change of Censorship Control.} Airports bypass the GFW, but many do not provide unrestricted Internet access. We find that most tested airports block access to certain websites, including Falun Gong-related sites, overseas news media, and other circumvention resources (\S\ref{sec:self-censorship}). This behavior may reflect operator risk management, but it may also reflect user demand. Some users may primarily want stable access to foreign entertainment and AI tools, and may tolerate operator-imposed filtering as long as the services they are interested in continue to work. Prior work on censorship attitudes similarly suggest that not all users conceptualize circumvention as completely unrestricted political access~\cite{participatoryCensorship2024,guo2012supportCensorship}, shedding light on why self-censorship is a trade-off they are willing to accept.

\parhead{Recommendations.}
The airport ecosystem suggests two
priorities for the circumvention community. 
First, build for operators,
not only users: 
the ecosystem's resilience flows from toolkits that let people inside China stand up and secure their own services, 
so hardening
and distributing that software---minimal-data account systems, 
safer defaults, 
a vulnerability-disclosure path for the shared templates that
95\% of airports depend on---may do more than building and running
yet another platform from outside the censored region. 
Closing that gap may matter more than any throughput gain. 
Second, evaluate circumvention by what users actually receive: bypassing the state censor guarantees neither service availability (\S\ref{sec:server-side-unblocking}) nor an uncensored path (\S\ref{sec:self-censorship}), 
so both belong alongside reachability and performance in how we measure success.

\bibliographystyle{IEEEtran}
\bibliography{censor,airport}

\crefalias{section}{appendix}
\appendices
\section{Ethical Considerations}
\label{sec:ethics}

This work documents a censorship circumvention ecosystem both operated and used by people who may face legal or personal risk for bypassing government controls. We considered the potential risks and benefits to individual airport operators and users, destination services, survey participants, the research team, and the airport ecosystem at large presented by conducting and publishing this research.

\parhead{Decision to Publish.} While airports have received little attention in the academic literature, they are not secret. Operators post public videos on YouTube describing how to run an airport (e.g.,~\cite{yt_v2board_tutorial,yt_sspanel_tutorial}), portal software is publicly available on Github (e.g.,~\cite{v2board, sspanel}), and public Telegram channels routinely advertise them.
Even operational details, such as airports' use of IEPL dedicated lines, are discussed openly in public forums and websites~\cite{V2EX2026IEPLAirportLinePull,NodeSeek2024ShenzhenHongKongIEPL,Ermaozi2025AirportLineType,OpenNetCN2026ClashAirportRoutes}.
Airports are regularly shut down, and airport operators have been targeted by the Chinese government (e.g.,~\cite{MuchuanProcuratorate2017, DongguanCourt2017, DuiHua2019VPNPartII,zhou2022nong_case,tan2026chinafears,ye2024seizing_innocence,zhou2024acquittal}), indicating that their existence is well known. 
Publishing this study could marginally help adversaries understand the ecosystem, but Chinese authorities likely know far more than we have been able to glean from public materials and remote measurements.
On the other hand, there is considerable opportunity for the academic research community to improve the airport ecosystem and build the next generation of circumvention tools that better protect 
users. 
Given this tradeoff, we argue that it is most beneficial to users to publish our investigation. In the remainder of the section, we discuss specific tradeoffs and how we limit risk to specific individuals and airports.

\parhead{Survey Participants.}
Our survey asked participants to describe experiences with tools that may be
legally sensitive. To reduce risk, we conducted the survey anonymously using a U.S.-based platform, did not
collect personally identifying information, allowed participants to skip any
question, and did not provide compensation that might pressure participation or would have required additional tracking. The survey and consent materials were reviewed and approved by the Institutional Review Board at the institution where the survey was conducted and analyzed. We do not release raw survey responses; the paper includes only aggregated results and short anonymized excerpts.

\parhead{Airport Services.}
Our active measurements used subscriptions we purchased through normal customer
channels and stayed within provider-enforced bandwidth limits. To reduce the
risk of disrupting service, throughput tests downloaded 50~MB objects and were
limited to a small purchased sample of airports. Our measurements may still have
imposed some load on airport infrastructure, but the traffic volume was designed
to resemble ordinary paid use. To prevent increased risk to any specific airport, we anonymized the specific airports we used and we are not publishing the fingerprints we developed for identifying airports.

\parhead{Censorship Measurements.}
Testing access to sensitive destinations can create risk if requests are issued
from uninvolved hosts or exposed to local network monitors. However, airports are specifically designed to enable access to censored content, and, as we show, block access to content they deem too risky. We conducted experiments from infrastructure registered under the name of a non-Chinese researcher who was aware of the associated risk. 

\parhead{Public Telegram Channels.}
Table~\ref{tab:channels} identifies seven public Telegram promotion channels by
handle. We weighed the reproducibility benefit against the risk of creating a
durable pointer for adversaries. We include the handles because these channels
are public, have more than 5,000~members each, are trivially discoverable through Telegram search using the keyword ``\chinese{机场},''. We do not identify individual operators or moderators, and we withhold the discovered airport portal list and scanning fingerprints.

\parhead{Researcher Safety.}
Purchasing subscriptions through payment channels such as Alipay carries risk because transactions are tied to an individual's identity, which could create legal or personal risk for researchers or collaborators with ties to China. When paid subscriptions were necessary, we used credentials belonging to a person who neither resides in nor is a citizen of China. Additionally, we created an email address used solely for this study to reduce the identifying information exposed to airport operators.

\section{Artifact Availability}
\label{sec:artifacts}

We will provide anonymized artifacts for review to the extent legally and
ethically possible. The shared artifacts include sanitized analysis code,
aggregate measurement outputs, and documentation sufficient to evaluate the
methodology. We withhold raw survey responses, Telegram histories, the list
of airport portals, purchased subscription credentials, and fingerprinting code
that would enable bulk discovery of airport infrastructure. These exclusions
protect 
the studied
ecosystem. The current anonymized artifact package is available at:
\url{https://anonymous.4open.science/r/airport-ecosystem-artifacts-616A}

\section{Generative AI Usage}
\label{sec:generative-ai}
Generative AI was used for editorial purposes in this manuscript, and all outputs were inspected by the authors to ensure accuracy and originality.

\section{Survey Details}
\label{appendix:survey}
Participants had the option of viewing the survey in Simplified Chinese (default) or
English. The following consent form and questions are presented as they appeared in the English version of the survey.

\parhead{Consent Form}
\label{appendix:consent-form}

\noindent DESCRIPTION: You are invited to participate in a research study on how
people circumvent China's Internet censorship system. We are a team of
anti-censorship researchers from
\ifthenelse{\boolean{anonymize-authors}}{[REDACTED]}{Stanford University, the
University of Colorado Boulder, and GFW Report}. Our goal is to better
understand and improve the circumvention ecosystem.  If you use VPNs, proxies,
and/or any other censorship circumvention tools to get around the Great Firewall
(GFW), we would greatly appreciate your help by filling out a short anonymous
survey. The survey will ask about your experiences and methods for bypassing
censorship.  Please note that the survey is completely anonymous, and no
personally identifiable information will be collected. The data collected will
be used to enhance our understanding and will contribute to developing more
effective circumvention tools. Our study has been reviewed and approved by the
\ifthenelse{\boolean{anonymize-authors}}{[REDACTED]}{Stanford University}
Institutional Review Board (IRB).

\noindent TIME INVOLVEMENT: Your participation will take approximately 5
minutes.

\noindent PAYMENTS: You will not receive payment for your participation.

\noindent RISKS AND BENEFITS: There is no foreseeable risk in this study. Study
data will be stored securely, in compliance with
\ifthenelse{\boolean{anonymize-authors}}{[REDACTED]}{Stanford University}
standards, minimizing the risk of confidentiality breach. The benefits which may
reasonably be expected to result from this study are a better understanding and
improvement of the circumvention ecosystem. We cannot and do not guarantee or
promise that you will receive any benefits from this study.

\noindent PARTICIPANT’S RIGHTS: If you have read this form and have decided to
participate in this project, please understand your participation is voluntary
and you have the right to withdraw your consent or discontinue participation at
any time without penalty or loss of benefits to which you are otherwise
entitled. The alternative is not to participate. You have the right to refuse to
answer particular questions. The results of this research study may be presented
at scientific or professional meetings or published in scientific journals. Your
individual privacy will be maintained in all published and written data
resulting from the study.

\parhead{Questions}
\label{appendix:survey-questions}

\noindent{1. How do you get censorship circumvention services these days?}
\begin{itemize}
    \item Purchase Commercial VPN (ExpressVPN, Astrill VPN, NordVPN, SurfsharkVPN, ProtonVPN, \begin{CJK}{UTF8}{gbsn}老王\end{CJK}VPN, etc.)
    \item Purchase ``Airport'' subscriptions
    \item I set up circumvention services myself
    \item My friend or family shared with me
    \item Psiphon3
    \item Lantern
    \item Tor
    \item Geph
    \item Cloudflare Warp 1.1.1.1
    \item FreeGate/UltraSurf
    \item I don't know
    \item Other (specify)
    \item Refuse to answer
\end{itemize}

\noindent{2. How much do you approximately pay to circumvent each month (in RMB)? (Enter -1 if you refuse to answer)}

\noindent{3. Where did you hear about the ``airports'' you've used?~\footnote{This question only appeared if the respondent had chosen `Purchase ``Airport'' subscriptions' in Question 1.}}
\begin{itemize}
    \item From family or friend
    \item Telegram Groups
    \item WeChat Group
    \item Internet Search
    \item I do not remember
    \item Other (specify)
    \item Refuse to answer
\end{itemize}

\noindent{4. Why do you choose to use the current circumvention tools?}
\begin{itemize}
    \item This is one of those available to me
    \item This is one of those I know how to use
    \item It is relatively cheaper
    \item It is relatively faster
    \item It is relatively more stable
    \item It is relatively easier to configure
    \item Other (specify)
    \item Refuse to answer
\end{itemize}

\noindent{5. How long does your proxy typically work uninterrupted in the last two years (before you have to purchase/install/configure a new one)?}
\begin{itemize}
    \item Up to 1 hour
    \item Less than 1 day
    \item Less than 1 week
    \item Less than 1 month
    \item Less than 3 months
    \item Less than 6 months
    \item Less than 9 months
    \item Less than 1 year
    \item More than 1 year
    \item I don't know
    \item Other (specify)
    \item Refuse to answer
\end{itemize}

\noindent{6. What protocols do you use to bypass censorship?}
\begin{itemize}
    \item I don't know
    \item Shadowsocks
    \item ShadowsocksR
    \item Hysteria2
    \item Naiveproxy
    \item Trojan
    \item VMess
    \item VLESS
    \item Wireguard
    \item TUIC
    \item Juicity
    \item SSH
    \item Snell
    \item Brook
    \item Socks5
    \item HTTP
    \item Others (please specify)
    \item Refuse to answer
\end{itemize}

\noindent{7. What are the challenges or incovenience you face with your current censorship circumvention setup? (Enter -1 if you refuse to answer)}

\section{Measured Airports}
\label{appendix:measured-airports}

\autoref{table:airport-information} details the characteristics of the airports in our dataset,
where an age marked `N/A' signifies that the airport was not found in the Telegram channels we analyzed.
\renewcommand{\arraystretch}{0.8}
\begin{table}[h]
\centering
\footnotesize
\caption{\textbf{Airports in our dataset and their characteristics---}\textnormal{Some airports with their age marked `N/A' may have been present on Telegram with a different TLD (e.g., example.com and example.xyz), but for consistency, we consider different domains to be different airports, even if they potentially have the same backend.}}
\label{table:airport-information}

\begin{minipage}[h]{0.48\linewidth}
\centering
\setlength{\tabcolsep}{3pt}
\begin{tabular}{l l l}
\toprule
\textbf{Airport} & \textbf{\$/TB} & \textbf{Date} \\
\midrule
Inexpensive 1   & 0.45  & 2022-03-06 \\
Inexpensive 2   & 0.44  & N/A        \\
Inexpensive 3   & 0.17  & N/A        \\
Midprice 1      & 10.3  & N/A        \\
Midprice 2      & 9.92  & N/A        \\
Midprice 3      & 10.7  & N/A        \\
Expensive 1     & 27.35 & 2022-02-28 \\
Expensive 2     & 12.11 & N/A        \\
Expensive 3     & 28.44 & 2023-01-30 \\
Scan 1          & 7.29  & N/A        \\
Scan 2          & 1.09  & N/A        \\
Scan 3          & 23.83 & N/A        \\
Scan 4          & 11.64 & N/A        \\
Scan 5          & 14.1  & N/A        \\
Scan 6          & 12.15 & N/A        \\
Scan 7          & 7.29  & N/A        \\
Scan 8          & 1.12  & N/A        \\
Scan 9          & 5.83  & N/A        \\
\bottomrule
\end{tabular}
\end{minipage}
\hfill
\begin{minipage}[h]{0.48\linewidth}
\centering
\setlength{\tabcolsep}{3pt}
\begin{tabular}{l l l}
\toprule
\textbf{Airport} & \textbf{\$/TB} & \textbf{Date} \\
\midrule
Young 1          & 15.8  & 2024-12-24 \\
Young 2          & 21.01 & 2024-09-24 \\
Young 3          & 11.4  & 2024-09-22 \\
Middle-aged 1    & 36.47 & 2022-04-05 \\
Middle-aged 2    & 21.01 & 2022-04-06 \\
Middle-aged 3    & 10.95 & 2022-03-31 \\
Old 1            & 14.59 & 2020-04-10 \\
Old 2            & 12.15 & 2019-12-18 \\
Old 3            & 21.88 & 2020-04-12 \\
Popular 1        & 26.8  & 2023-10-24 \\
Popular 2        & 21.15 & 2022-02-24 \\
Popular 3        & 15.75 & N/A        \\
Popular 4        & 16.48 & 2024-01-10 \\
Popular 5        & 5.11  & N/A        \\
Popular 6        & 0.97  & 2022-02-25 \\
Popular 7        & 7.24  & N/A        \\
Popular 8        & 12.16 & 2022-02-25 \\
\bottomrule
\end{tabular}
\end{minipage}

\end{table}

\end{document}